\documentclass[aps,prb,amsmath,amssymb,floatfix,twocolumn,superscriptaddress,10pt]{revtex4-1}
\usepackage{graphicx}
\usepackage{subfigure}
\usepackage{hyperref}
\usepackage{braket}
\usepackage{bm}

\DeclareMathAlphabet\mathbfcal{OMS}{cmsy}{b}{n}

\def\YBCOs{YBa$_2$Cu$_3$O$_{6+x}$ }

\def\YBCO248{YBa$_2$Cu$_4$O$_8$}

\def\HBCOs{HgBa$_2$CuO$_{4+\delta}$ }

\def\BSCCOs{Bi$_2$Sr$_2$CaCu$_2$O$_{8+x}$ }
\def\BSCCOns{Bi$_2$Sr$_2$CaCu$_2$O$_{8+x}$}

\def\etal{{\it et al.}}

\def\bnabla{\bm{\nabla}}

\def\br{{\bf r}}
\def\ba{{\bf a}}
\def\bq{{\bf q}}
\def\bp{{\bf p}}
\def\vs{{\bm v}_s}
\def\bA{{\bf A}}
\def\bB{{\bf B}}
\def\cbA{{\mathbfcal{A}}}
\def\bR{{\bf R}}
\def\bk{{\bf k}}
\def\bq{{\bf q}}
\def\bQ{{\bf Q}}

\def\prl{{Phys. Rev. Lett. }}
\def\prb{{Phys. Rev. B }}
\def\rmp{{Rev. Mod. Phys. }}

\def\natcomm{{Nat. Commun. }}

\begin{document}

\title{Quantum oscillations from a pair-density wave}

\author{Yosef~Caplan}
\affiliation{Racah Institute of Physics, The Hebrew University,
  Jerusalem 91904, Israel}
\author{Dror~Orgad}
\affiliation{Racah Institute of Physics, The Hebrew University,
  Jerusalem 91904, Israel}

\date{\today}

\begin{abstract}

A pair-density wave state has been suggested to exist in underdoped cuprate superconductors,
with some supporting experimental evidence emerging over the past few years from scanning tunneling spectroscopy.
Several studies have also linked the observed quantum oscillations in these systems to a reconstruction
of the Fermi surface by a pair-density wave. Here, we show, using semiclassical analysis and numerical
calculations, that a Fermi pocket created by first-order scattering from a pair-density wave cannot induce such
oscillations. In contrast, pockets resulting from second-order scattering can cause oscillations. We consider
the effects of a finite pair-density wave correlation length on the signal, and demonstrate that it is only weakly sensitive
to disorder in the form of $\pi$-phase slips. Finally, we discuss our results in the context of
the cuprates and show that a bidirectional pair-density wave may produce observed oscillation frequencies.

\end{abstract}

\maketitle

\section{Introduction}

The underdoped cuprate high-temperature superconductors are known to harbor a variety of electronic
orders.\cite{intertwined} Among them, charge-density waves (CDW) have attracted attention over recent years owing 
to a series of experimental observations.\cite{CDW-review}
Much more illusive is the pair-density wave (PDW) that is associated with a spatially oscillating superconducting
order parameter of zero mean.\cite{PDW-review} Such a state has its origin in the superconducting FFLO phase,\cite{FF,LO}
which can emerge in a magnetic field. Subsequently, the PDW was conjectured to occur without explicit time-reversal breaking
in the cuprates.\cite{PDW-Himeda,PDW-Berg,PDW-Lee} Numerical studies suggest that the PDW may be energetically close to the uniform
superconducting state,\cite{PDW-Himeda,PDW-2d-tJ,PDW-Yang,Kopp,Corboz} and recent scanning tunneling spectroscopy gives evidence that
it is realized within halos surrounding Abrikosov vortex cores in \BSCCOns.\cite{Halo-Edkins}

Electronic orders that break translational symmetry naturally lead to redistribution of spectral weight in momentum space.
If the latter contains gapless Fermi segments, as is the case in the underdoped cuprates, this can result in the formation
of Fermi pockets, which give rise to quantum oscillations. It is commonly believed that the observed quantum oscillations in
the high-temperature superconductors are due to such a scenario.\cite{Sebastian-review} In particular, the established presence
of a bidirectional CDW in these materials offers a natural route for formation of electron-like pockets with the required
area to match the oscillations' frequency.\cite{H-S,Sachdev} Nevertheless, the fact that the correlation length of the bidirectional CDW
is shorter than the cyclotron radius of the closed orbit responsible for the oscillations casts doubt on this picture.\cite{Gannot}
Other options, including unidirectional CDW,\cite{Norman,Steve-uni,Mohit} and coexisting $d$-wave superconductivity with 
$d$-density wave,\cite{DDW-Lee,DDW-Chakravarty} or loop current order,\cite{loop-Senthil,loop-Vafek} were also considered.

A PDW is appealing from the perspective of generating quantum oscillations in that it provides a superconducting state with
gapless Fermi arcs.\cite{Shirit} This fact has led to suggestions that a PDW is the source of the oscillations in
the pseudogap regime.\cite{Zelli-QO,Davis-Norman,PDW-Senthil} However, these studies differ in their reconstruction schemes.
Refs. \onlinecite{Zelli-QO,Davis-Norman} considered pockets, which we dub first-order pockets, that are generated by first-order
scattering and comprised of Fermi segments shifted by the PDW wave vector ${\bf Q}_x$
(or ${\bf Q}_y$ in the case of bidirectional order). Yet, their ability to produce oscillations was subsequently
questioned by Ref. \onlinecite{PDW-Senthil} on the basis that they include both electron-like and hole-like pieces.
Instead, other pockets, designated by us as being second order, were suggested. These are constructed
from $2{\bf Q}_x$ and $2{\bf Q}_y$ translated Fermi segments, all having the same character.

Motivated by the above disagreement, the lack of detailed and approximation-free calculations, and by the experimental situation
we carry out a theoretical investigation of quantum oscillations from a PDW. We show, using both semiclassical analysis and
numerical calculations, that the coupling of the Bogoliubov quasiparticles to the superfluid velocity field around vortices,
$\vs(\br)$, plays a pivotal role in establishing the structure of the local density of states (DOS) and hence of the oscillations spectrum.
If one ignores this coupling then every Fermi pocket, regardless of its order, supports a periodic semiclassical orbit along constant
energy contours $E(\bk)$, where $\bk$ is the gauge invariant crystal momentum. However, in the presence of the coupling
semiclassical motion takes place along constant $E(\bk\pm m\vs/\hbar)$ electron-like and hole-like segments, respectively.
Hence, first-order pockets that contain both types of segments generally do not sustain closed momentum-space orbits in the presence of
a magnetic field. Moreover, in real space the motion of electrons and holes occurs under the influence of opposite effective
magnetic fields, further hindering the formation of periodic orbits. In contrast, a semiclassical wave packet maintains its
character around a second-order pocket and is able to execute periodic motion that is largely free of these problems. The 
outcome is similar to the one arising from a $2\bQ_{x,y}$ CDW, except for the electron-hole mixing that is
present near PDW scattering points. This mixing leads to broadened oscillations     
at a frequency that nearly obeys the familiar Onsager relation to the area of the pocket.
%
%

The effect of PDW phase disorder can be analysed semiclassically along the lines of Ref. \onlinecite{Gannot}. It takes the
form of a Berry phase, to which every Bragg scattering event off the PDW contributes the local PDW phase.
The result is a generic suppression of the oscillatory signal by a Dingle factor that decays exponentially with the inverse PDW
correlation length. However, we note an interesting possibility that arises from the fact that the oscillatory DOS originates from
second-order pockets, where each scattering between Fermi segments contributes twice the local PDW phase. Hence, disordering
of the PDW by $\pi$-phase slips leads to a Berry phase, which is a multiple of $2\pi$ and has no effect on the quantization.
We show that while such complete immunity is not exact the DOS is indeed much less sensitive to disorder of this type.

The next three sections are devoted to substantiating the above results by defining a model Hamiltonian, analyzing it semiclassically
and diagonalizing it numerically. We then discuss the relevance of our findings to the cuprates. We argue that it is possible that
the quantum oscillations in \YBCOs and \HBCOs are due to a yet undiscovered PDW that is bidirectional
with half the wave vector of the CDW that was detected in these systems at low magnetic fields. We also calculate the
expected oscillation frequency from a PDW of the nature suggested by the observations in \BSCCOns.

\section{Model}

We consider a model of a bidirectional $d$-wave PDW residing on the bonds of a square lattice with unit lattice constant and size $L\times L$.
The PDW wave vectors in the $\alpha=x,y$ directions are taken to be commensurate $Q_\alpha=2\pi/\lambda_\alpha$, with integer
$\lambda_\alpha$. We allow for independent positional disorder in the $x$ and $y$ PDW components (via phases $\theta^x$ and $\theta^y$),
but assume that they share the same superconducting phase, $\phi$, which is disordered by common vortices. We neglect PDW amplitude
fluctuations due to vortices or otherwise. Incorporating the effects of a transverse magnetic field, $\bB=B\hat z$, by a Peierls substitution
we are led to study the following Bogoliubov-de Gennes (BdG) Hamiltonian
\begin{eqnarray}
\label{eq:H1}
\nonumber
{\cal H}&=&\sum_\br\sum_{\ba=\pm\hat{x},\pm\hat{y}}\Psi^\dagger_{\br+\ba}
\left(
\begin{array}{cc}
h_{\br,\ba}(A) & \Delta_{\br,\ba}(\phi) \\
\Delta^*_{\br,\ba}(\phi) & -h^*_{\br,\ba}(A)
\end{array}\right)\Psi_\br \\
&&-\mu\sum_\br\Psi^\dagger_\br\tau_3 \Psi_\br,
\end{eqnarray}
where $\Psi^\dagger_\br=(c_{\br\uparrow}^\dagger,c_{\br\downarrow})$, ${\bm \tau}$ are the Pauli matrices, and
\begin{eqnarray}
\label{eq:hdef}
&&h_{\br,\ba}(A)=-t e^{-i A_{\br, \ba}}, \\
\nonumber
\label{eq:Deltadef}
&&\Delta_{\br,\ba}(\phi)=\sum_{\alpha=x,y}\Delta_\ba^\alpha\cos[\bQ_\alpha\cdot(\br+\ba/2)+\theta^\alpha_{\br+\ba/2}]
e^{i\phi_{\br+\ba/2}}.\\
\end{eqnarray}
Here, $A_{\br,\ba}=(e/\hbar c)\bA_{\br+\ba/2}\cdot\ba$, and the $d$-wave form factor is given by $\Delta^\alpha_{\pm\hat{x}}=\Delta^\alpha$,
$\Delta^\alpha_{\pm\hat{y}}=-\Delta^\alpha$. For the phase field we use the decomposition
$\phi_{\br+\ba/2}= \phi_\br+{\bm \nabla}\phi_{\br+\ba/2}\cdot \ba/2\equiv \phi_\br+\nabla\phi_{\br,\ba}/2$, where the site field $\phi_\br$
is defined mod $2\pi$ but the bond gradient field ${\bm \nabla}\phi_{\br+\ba/2}$ is single valued. As a result the phase factor
in Eq. (\ref{eq:Deltadef}) is well defined and one also finds that $\phi_{\br+\ba}=\phi_\br+\nabla\phi_{\br,\ba}$ mod $2\pi$. A similar
decomposition is used for $\theta^\alpha$.

Next, we remove the site variables $\phi_\br$ from ${\cal H}$ by the single valued unitary transformation,\cite{Anderson}
\begin{equation}
\label{eq:Anderson}
\Psi_\br\rightarrow\left(\begin{array}{cc} 1& 0 \\ 0 & e^{-i\phi_\br} \end{array}\right) \Psi_\br,
\end{equation}
and introduce the superfluid velocity field ${{\vs}_{}}_{\br+\ba/2}=(\hbar/2m)[{\bm \nabla}\phi_{\br+\ba/2}+(2e/\hbar c)\bA_{\br+\ba/2}]$.
Consequently, in terms of ${{v_s}_{}}_{\br,\ba}=(m/\hbar) {{\vs}_{}}_{\br+\ba/2}\cdot \ba$, the Hamiltonian becomes
\begin{eqnarray}
\label{eq:H2}
\nonumber
{\cal H}&=&\sum_{\br,\ba}e^{i({{v_s}_{}}_{\br,\ba}-A_{\br,\ba})}\Psi^\dagger_{\br+\ba}
\left(
\begin{array}{cc}
h_{\br,\ba}(v_s) & \Delta_{\br,\ba}(0) \\
\Delta^*_{\br,\ba}(0) & -h^*_{\br,\ba}(v_s)
\end{array}\right)\Psi_\br \\
&&-\mu\sum_\br\Psi^\dagger_\br\tau_3 \Psi_\br,
\end{eqnarray}
where $h_{\br,\ba}(v_s)$ and $\Delta_{\br,\ba}(0)$ are defined by the appropriate substitutions into
Eqs. (\ref{eq:hdef}) and (\ref{eq:Deltadef}).

\subsection{Solution for constant fields}

Anticipating the semiclassical treatment of the next section, we are interested in the Bloch states of the above Hamiltonian
when the "external fields" take constant values, i.e., $\bA_{\br+\ba/2}=\bA$, ${{\vs}_{}}_{\br+\ba/2}=\vs$, and
$\theta_{\br+\ba/2}^\alpha=\theta^\alpha$. These states serve us in order to construct a wave packet whose motion
we analyze in the case where the fields vary slowly in space.
Given the translational symmetry of the problem at hand we expand
\begin{equation}
\label{eq:kexpand}
\Psi_\br=L^{-1}e^{i(\frac{m}{\hbar}\vs-\frac{e}{\hbar c}\bA)\cdot\br}\sum_\bk\sum_\bq e^{i(\bk+\bq)\cdot \br}\Psi_{\bk+\bq},
\end{equation}
where $\Psi^\dagger_\bk=(c_{\bk\uparrow}^\dagger,c_{-\bk\downarrow})$. The sum over the gauge invariant crystal momentum $\bk=(2\pi/L)(n_x\hat{x}+n_y\hat{y})$, runs over the reduced Brillouin zone (RBZ), $n_\alpha=-(N_\alpha-1)/2,\cdots, (N_\alpha-1)/2$, with $N_\alpha=L/\lambda_\alpha$ the number of unit cells in the $\alpha$ direction. The $\Lambda=\lambda_x\lambda_y$ Bragg vectors are
$\bq=m_x\bQ_x+m_y\bQ_y$ with $m_\alpha=-(\lambda_\alpha-1)/2,\cdots, (\lambda_\alpha-1)/2$. Expressed in momentum space the Hamiltonian reads
${\cal H}=\sum_\bk {\cal H}(\bk)=\sum_\bk\sum_{\bq,\bq'}\Psi^\dagger_{\bk+\bq}H_{\bq\bq'}(\bk)\Psi_{\bk+\bq'}$, where
\begin{eqnarray}
\label{eq:Hk}
\nonumber
H_{\bq\bq'}(\bk)&=&\left(\begin{array}{cc} \xi_{\bk+\bq+\frac{m}{\hbar}\vs} & 0 \\ 0 & -\xi_{\bk+\bq-\frac{m}{\hbar}\vs}
\end{array}\right)\delta_{\bq,\bq'} \\
\nonumber
&&+\sum_\alpha\left(e^{i\theta^\alpha}\Delta_{\bk+\bq-\frac{\bQ_\alpha}{2}}^\alpha\delta_{\bq',\bq-\bQ_\alpha}\right. \\
&&\hspace{0.79cm}\left. +e^{-i\theta^\alpha}\Delta_{\bk+\bq+\frac{\bQ_\alpha}{2}}^\alpha\delta_{\bq',\bq+\bQ_\alpha}\right)\tau_1,
\end{eqnarray}
$\xi_\bk=-2t(\cos k_x + \cos k_y)-\mu$, and $\Delta^\alpha_\bk = \Delta^\alpha(\cos k_x - \cos k_y)$.

The spectrum of ${\cal H}$ is discussed in Appendix \ref{App:diag}. There we show that in the presence of $\vs$
it consists of doubly degenerate bands whose number, $\Lambda_\bk$, may change over the RBZ but sums up to the total number
of degrees of freedom, owing to $\Lambda_{-\bk}=\Lambda-\Lambda_\bk$. The excitations are created from the ground state, $|g\rangle$,
by the quasiparticle operators $\gamma^\dagger_{\bk n\pm}$, where $n$ is the band index, and take the form
\begin{equation}
\label{eq:qpmain}
\gamma^\dagger_{\bk n\pm}|g\rangle=\frac{\Lambda^{1/2}}{L}
\sum_\br e^{i(\bk\pm\frac{m}{\hbar}\vs \mp\frac{e}{\hbar c}\bA)\cdot\br}\left\{
\begin{array}{c} \Psi_\br^\dagger\varphi_{\bk n+}(\br)|g\rangle \\
\varphi^\dagger_{\bk n-}(\br)\Psi_\br |g\rangle \end{array}.\right.
\end{equation}
The periodic parts, $\varphi_{\bk n\pm}$, of the Bloch states are given in Eq.~(\ref{eq:phis}) and depend on
$\vs$ and $\theta^\alpha$.

\section{Semiclassical analysis}

\subsection{Wave packet}
\label{sub:wavepacket}

Our goal is to obtain the DOS by semiclassical quantization of periodic cyclotron orbits. To this end, we follow Ref. \onlinecite{Niu-semi}
and consider a quasiparticle wave packet that is centered around $\bk_c$ in momentum space and constructed from Bloch states,
Eq. (\ref{eq:qpmain}), of the "local Hamiltonian" ${\cal H}_c={\cal H}[\bA(\br_c),\vs(\br_c),\theta^\alpha(\br_c)]$
\begin{eqnarray}
\label{eq:wavepacket}
\nonumber
|\Psi_{n\eta}(\bk_c,\br_c)\rangle&=&\sum_\bk W(\bk-\bk_c)e^{-i\eta\bk\cdot[\br_c-{\cbA_k}_{n\eta}(\bk_c,\br_c)]}\\
&\times&\gamma^\dagger_{\bk n\eta}(\br_c)|g(\br_c)\rangle,
\end{eqnarray}
with $\eta=\pm$. Here, $W(\bk)$ is a real weighting function that is peaked around $\bk=0$, with width much smaller than $\bQ_{x,y}$
and the typical wave vector of variations in the external fields.
It is also assumed to be periodic over the RBZ and square-normalized $\sum_\bk W^2(\bk)=1$. The $\bk$-space Berry connection is given by
\begin{equation}
\label{eq:Berryk}
\cbA_{k_{n\eta}}(\bk_c,\br_c)=i\sum_\bk W^2(\bk-\bk_c)\langle\varphi_{\bk n\eta}|\bnabla_\bk\varphi_{\bk n\eta}\rangle,
\end{equation}
with its $\br_c$ dependence inherited from the dependence of $\varphi$ on $\bA$,$\vs$ and $\theta^\alpha$,
and where the overlap is defined by a sum over a unit cell in real space
\begin{equation}
\label{eq:defbrak}
\langle\varphi_{\bk n\eta}|\bnabla_\bk\varphi_{\bk n\eta}\rangle\equiv\sum_{\br\in {\rm u.c.}}\varphi^\dagger_{\bk n\eta}(\br)
\bnabla_\bk \varphi_{\bk n\eta}(\br).
\end{equation}
We note that Eq. (\ref{eq:phis}) implies $\varphi_{\bk n-}=i\tau_2\varphi_{\bk n+}^*$, which
can be used to show $\cbA_{k_{n+}}=\cbA_{k_{n-}}$.
Henceforth, we assume that the motion takes place within a single band and suppress the band indices.

Our choice for the phase of the wave packet in Eq. (\ref{eq:wavepacket}) is tacitly related to the identity of its position
in real space. While there is no ambiguity with regards to this question in the case of normal electronic systems the issue
is less clear for superconductors.\cite{Liang-semi,Niu2020} Most generally, one may consider a position operator of the form
$\hat\br=\sum_\br\br\Psi_\br^\dagger R \Psi_\br$, where $R$ is a hermitian matrix, and require that
$:\!\!\langle\Psi|\hat\br|\Psi\rangle\!\!:\,\equiv\langle\Psi|\hat\br|\Psi\rangle-\langle g|\hat\br|g \rangle=\br_c$.
For $R=\tau_3$ the operator is related to
the center of mass (or charge), while $R=I$ corresponds to the spin center. Because the charge of a superconducting quasiparticle
may change along its travel in momentum space, whereas its spin is conserved (barring spin-orbit coupling),
it was suggested that the spin center should be identified with the center of the wave packet.\cite{Niu2020}

Here, we would like to offer another argument in favor of using $R=I$. In order to achieve self-consistency of the semiclassical
treatment, the wave packet (\ref{eq:wavepacket}) needs to be concentrated in real space such that the local Hamiltonian
constitutes an approximate generator of its dynamics. Ideally, this would mean that for ${\cal H}=\sum_\br {\cal H}_\br[F(\br)]$,
where $F(r)$ represents the collection of external fields, $:\!\!\langle\Psi|{\cal H}|\Psi\rangle\!\!:\,=\,
:\!\!\langle\Psi|\sum_\br {\cal H}_\br[F(\br_c)] |\Psi\rangle\!\!:$, at least when ${\cal H}_\br$ is expanded to first order in $\br-\br_c$.
We note from Eq. (\ref{eq:H2}) that when $\theta^\alpha$ is constant
$\partial {\cal H}_\br/\partial\br_c\propto \sum_\ba \Psi^\dagger_{\br+\ba}\Psi_\br$. Thus, the above requirement is fulfilled in
the absence of PDW phase disorder if $:\!\!\langle\Psi|\hat\br|\Psi\rangle\!\!:\,=\br_c$, for $R=I$. We demonstrate
in Appendix \ref{App:expect} that this expectation value holds for wave packet (\ref{eq:wavepacket}).

\subsection{Dynamics}

The semiclassical dynamics of the coordinates $\bk_c$ and $\br_c$ is derived from the Lagrangian
$L=:\!\!\langle\Psi_\eta | i\hbar\frac{d}{dt}-{\cal H}_c|\Psi_\eta\rangle\!\!:$, whose calculation is outlined in Appendix \ref{App:expect}.
The result is
\begin{eqnarray}
\label{eq:L}
\nonumber
L&=&\dot{\br}_c\left[\eta\hbar\bk_c-\frac{e}{c}\tilde{\bA}(\br_c)\right]-E(\bk_c,\br_c)\\
&+&\eta\hbar\left[\dot{\bk}_c\cdot \cbA_k(\bk_c,\br_c)+\dot{\br}_c\cdot \cbA_r(\bk_c,\br_c)\right],
\end{eqnarray}
where
\begin{equation}
\label{eq:tildeA}
\tilde{\bA}(\br_c)=\bA(\br_c)-\frac{mc}{e}\vs(\br_c),
\end{equation}
and the second Berry connection
\begin{equation}
\label{eq:Berryr}
\cbA_r(\bk_c,\br_c)=i\sum_\bk W^2(\bk-\bk_c)\langle\varphi_{\bk}|\bnabla_{\br_c}\varphi_\bk\rangle,
\end{equation}
is also independent of $\eta$.

The ensuing Euler-Lagrange equations read
\begin{eqnarray}
\label{eq:ELr}
\dot{\br}_c&=&\frac{\eta}{\hbar}\bnabla_{\bk_c}E-\Omega_{kr}\dot{\br}_c-\Omega_{kk}\dot{\bk}_c,\\
\label{eq:ELk}
\dot{\bk}_c&=&-\frac{\eta}{\hbar}\bnabla_{\br_c}E-\frac{\eta e}{\hbar c}\dot{\br}_c\times\tilde{\bB}+\Omega_{rk}\dot{\br}_c+\Omega_{rr}\dot{\bk}_c,
\end{eqnarray}
where $\tilde{\bB}=\bnabla_{\br_c}\times\tilde{\bA}$ and the Berry curvatures are given by
\begin{equation}
\label{eq:Berrycurv}
\left(\Omega_{ab}\right)_{\alpha\beta}=\frac{\partial\left({\cal A}_b\right)_\beta}{\partial \left(a_c\right)_\alpha}
-\frac{\partial\left({\cal A}_a\right)_\alpha}{\partial \left(b_c\right)_\beta},
\end{equation}
with $a,b=\{k,r\}$ and $\alpha,\beta=\{x,y\}$.

\subsection{First-order scattering}

Quantum oscillations originate from the alternating presence and absence of states at the chemical potential, as the magnetic field is varied.
For a weak PDW, which is our focus here, such states correspond to periodic solutions of Eqs. (\ref{eq:ELr}) and (\ref{eq:ELk}),
for which the motion is largely along sections of the unperturbed Fermi surface. On these sections the eigenstates are either
"electron-like" with character close to $c^\dagger_{\eta\bk,\eta\uparrow}|g\rangle$, and energy $\xi(\bk_c+m\vs/\hbar)$,
or "hole-like" given approximately by $c_{-\eta\bk,-\eta\uparrow}|g\rangle$ with energy $-\xi(\bk_c-m\vs/\hbar)$, see Eq. (\ref{eq:Hk}).
The PDW may cause scattering of the wave packet between Fermi segments at points that are connected by an integer combination
$m_x\bQ_x+m_y\bQ_y$, where $m_x+m_y$ is the order at which the process appears in perturbation theory. Odd-order scattering
connects segments of opposite character, while even-order scattering preserves the nature of the segment.
Refs. \onlinecite{Zelli-QO,Davis-Norman} claimed to find periodic orbits that emerge from first-order scattering.
We proceed to show that this claim is erroneous and is caused by neglecting the effects of the superfluid velocity
on the motion of quasiparticles.

\begin{figure}[t!!!]
  \centering
  \includegraphics[width=0.93\linewidth,clip=true]{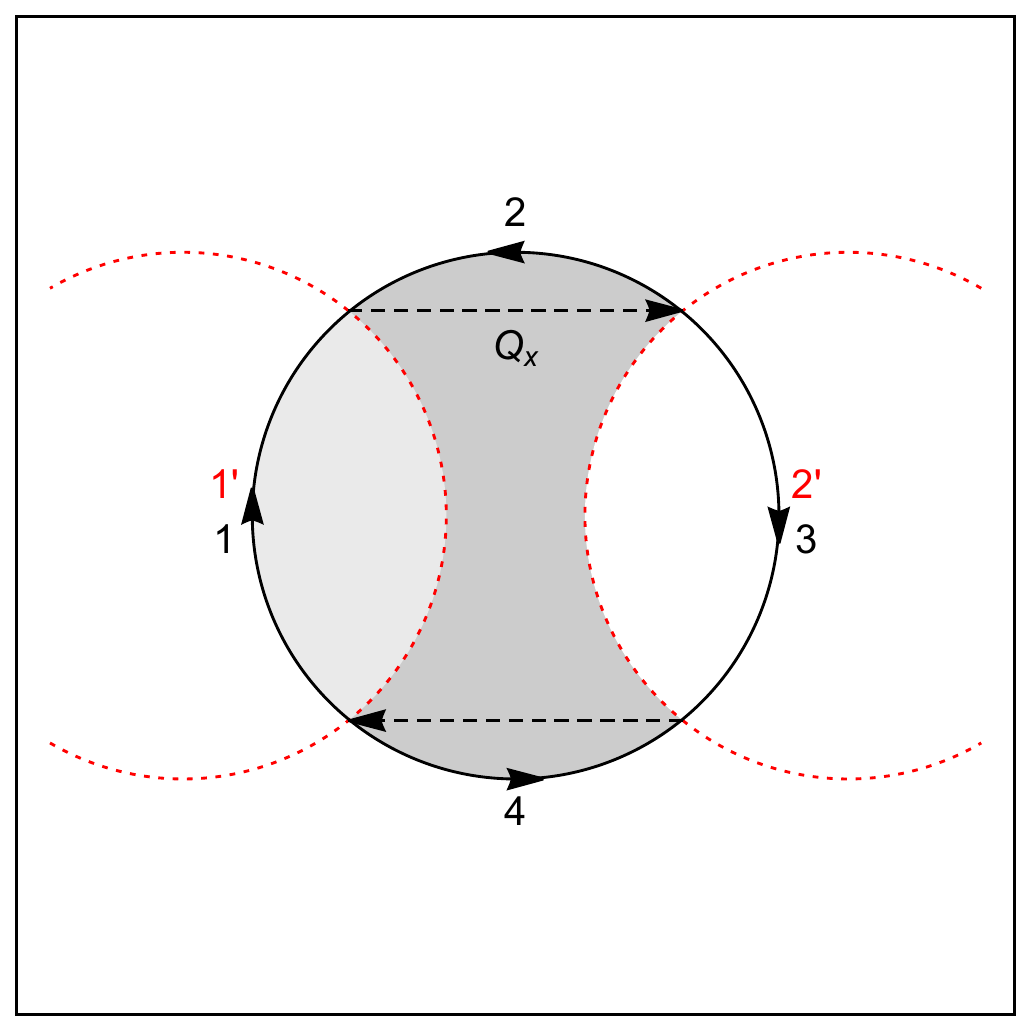}
  \caption{First-order scattering from a unidirectional PDW. If one neglects the superfluid velocity then periodic semiclassical motion
  occurs in $\bk_c$-space along segments $1\rightarrow 2\rightarrow 3\rightarrow 4$ of the unperturbed Fermi surface. The sense of rotation
  is reversed through each of the four Bragg reflections involved. As a result a (dark grey) pocket appears, constructed from shifted
  Fermi sections. Motion in $\br_c$-space proceeds along a rotated and scaled version of the $\bk_c$ orbit. In contrast, when accounting
  for the effects of $\vs$
  the sense of rotation is the same on all segments and one may conclude that a (light grey) pocket emerges due to $1'\rightarrow 2'$
  orbit. However, the picture is misleading since the two segments
  are drawn as function of two different parameters: $\bk_c\pm m\vs/\hbar$.
  Moreover, such a pocket would produce an open skipping trajectory for $\br_c$. }
  \label{fig:pocket1}
\end{figure}

To this end, we consider the semiclassical dynamics away from the scattering points in an ordered system $(\theta^\alpha=0)$,
where the Berry curvatures vanish. On an electron-like segment the energy is a function, $E(\bk_e)$, of the variable $\bk_e=\bk_c+m\vs/\hbar$.
Consequently, Eqs. (\ref{eq:ELr}) and (\ref{eq:ELk}) may be cast into the form
\begin{eqnarray}
\label{eq:esegk}
\dot{\bk}_e&=&-\frac{e}{\hbar^2 c}\bnabla_{\bk_e}\xi(\bk_e)\times \bB_{e\eta},\\
\label{eq:esegr}
\dot{\br}_c&=&\eta\frac{\hbar c}{e B_{e\eta}^2}\dot{\bk}_e\times \bB_{e\eta},
\end{eqnarray}
with the effective magnetic field $\bB_{e\eta}=\tilde\bB+\eta(mc/e)\bnabla_{\br_c}\times\vs$. For $\vs$
that is generated by a collection of vortices at positions $\bR_n$ one finds
$\bB_{e\eta}=\eta\bB+(1-\eta)(hc/2eB)\sum_n\delta(\br_c-\bR_n)\bB$.
Note that the equations imply a reversed $\br_c$ motion for down-spin excitations ($\eta=-$) as compared
to up-spin excitations ($\eta=+$). This is consistent with the fact that for down electrons $\br_c$ is the inverted center of mass
position.

Conversely, on a hole-like segment the energy is a function of the variable $\bk_h=\bk_c-m\vs/\hbar$ and the equations of
motion become
\begin{eqnarray}
\label{eq:hsegk}
\dot{\bk}_h&=&\frac{e}{\hbar^2 c}\bnabla_{\bk_h}\xi(\bk_h)\times \bB_{h\eta},\\
\label{eq:hsegr}
\dot{\br}_c&=&\eta\frac{\hbar c}{e B_{h\eta}^2}\dot{\bk}_h\times \bB_{h\eta},
\end{eqnarray}
with $\bB_{h\eta}=\tilde\bB-\eta(mc/e)\bnabla_{\br_c}\times\vs$. For a collection of vortices $\bB_{h\eta}=-\eta\bB+(1+\eta)(hc/2eB)\sum_n\delta(\br_c-\bR_n)\bB$.

If one follows Ref. \onlinecite{Zelli-QO} and sets $\vs=0$ then $\bk_e=\bk_h=\bk_c$ and $\bB_{e\eta}=\bB_{h\eta}=\bB$.
As a result, Eqs. (\ref{eq:esegk}) and (\ref{eq:hsegk}) imply that motion in $\bk_c$-space takes place on constant $\xi(\bk_c)$
contours, or more generally, according to Eq. (\ref{eq:ELk}) along constant $E(\bk_c)$ contours. The equations also indicate that the sense
of rotation is changed across each Bragg scattering. Assuming that $B$ is small enough to disregard magnetic breakdown
this condition determines the shape of the consequent pockets, see Fig. \ref{fig:pocket1}.
The latter coincide with the pockets that emerge from diagonalizing ${\cal H}$ for $\bB=\vs=0$.
Concomitantly, Eqs. (\ref{eq:esegr}) and (\ref{eq:hsegr}) tell us that $\br_c$ also executes periodic motion that is derived from that of
$\bk_c$ by a $\pi/2$ rotation and scaling by $l_B^2=\hbar c/eB$. Upon semiclassical quantization these periodic orbits would give rise
to Landau levels and hence to oscillations. However, this is an artefact of the approximation $\vs=0$.

In the presence of $\vs$, and as long as the semiclassical trajectory misses the vortex cores, $\bB_{e\eta}=-\bB_{h\eta}$. Thus,
motion on electron-like and hole-like segments follows constant $\xi(\bk_e)$ and $\xi(\bk_h)$ contours, respectively, with the
same sense of rotation for both cases. Superficially, this may lead to the conclusion that closed $\bk$-space pockets may
still form, as suggested by the light gray pocket in Fig. \ref{fig:pocket1}. However, it should be noted that the constancy
of the $\xi$ contours is defined relative to two different variables: $\bk_c\pm m\vs(\br_c)/\hbar$. Therefore, generically,
one does not expect that solving the coupled equations for $\bk_c$ and $\br_c$ would yield a closed orbit. Even if a closed
cycle is formed for $\bk_c$, the fact that $\bB_{e\eta}=-\bB_{h\eta}$ would lead by Eqs. (\ref{eq:esegr}) and (\ref{eq:hsegr}) to
an open skipping orbit for $\br_c$. Indeed, as we demonstrate by numerical calculations below, there is no evidence for
quantum oscillation from first-order scattering off a PDW.

\subsection{Second-order pockets}
\label{section:so}

\begin{figure}[t!!!]
  \centering
  \includegraphics[width=0.93\linewidth,clip=true]{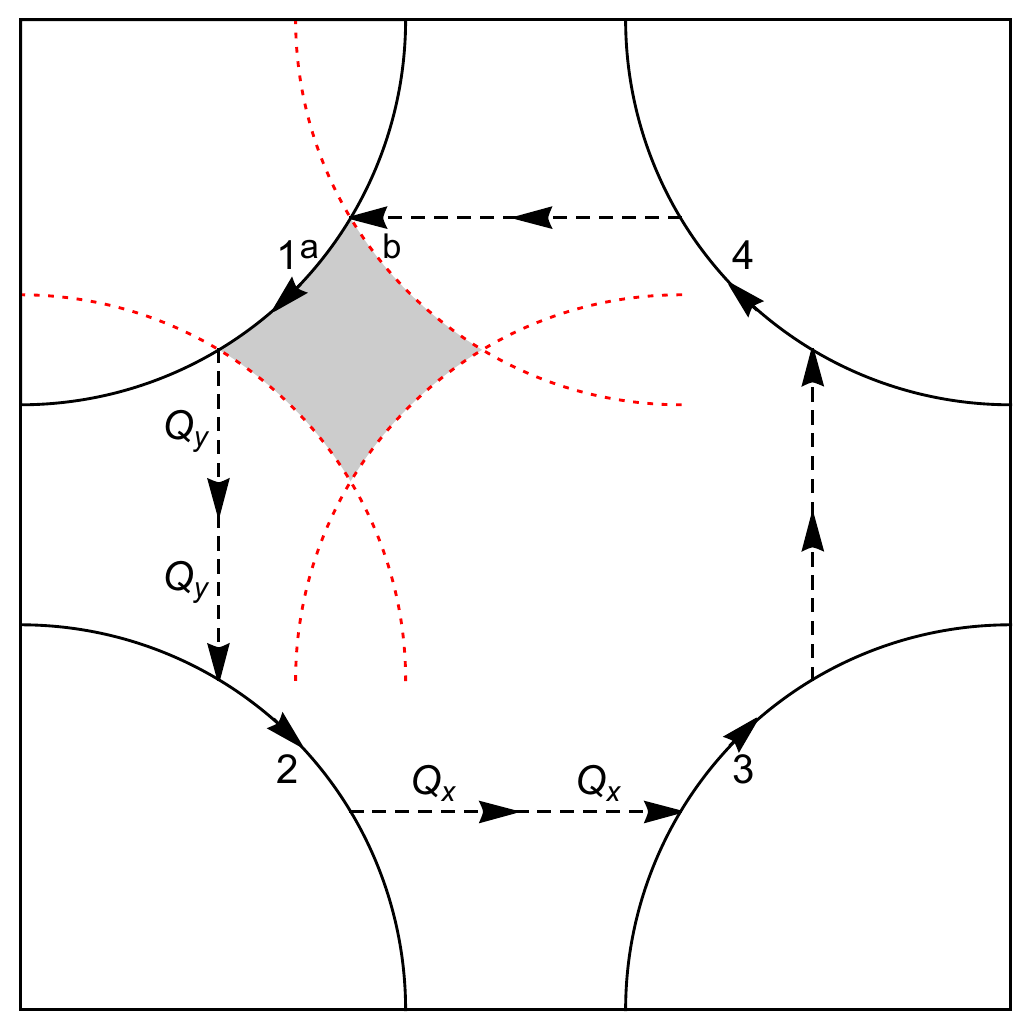}
  \caption{A second-order pocket from a bidirectional PDW. The semiclassical cycle $1\rightarrow 2\rightarrow 3\rightarrow 4$
  is for an up-spin electron-like excitation and is depicted in $\bk_e=\bk_c+m\vs/\hbar$ space. It consists of constant $\xi(\bk_e)$
  Fermi segments connected by second-order Bragg scattering via an intermediate state.}
  \label{fig:pocket2}
\end{figure}

In light of the preceding discussion we are led to consider the possibility of orbits that maintain their
character during semiclassical evolution. Under such circumstances we may expect the entire motion to unfold
as function of a single momentum variable, $\bk_e$ or $\bk_h$, and the real-space trajectory to evolve under
a unique effective magnetic field. We will show that this expectation is partially borne out by the following analysis.
Since each scattering off the PDW changes the identity of the orbit, it is necessary to construct it from Fermi segments
connected by second (or in general even) order scattering. Here we will do so for a bidirectional PDW, see Fig. \ref{fig:pocket2},
although it can also be done for a unidirectional PDW.

For the remaining discussion, let us focus on an electron-like excitation. Away from the scattering points and
for weak PDW the semiclassical dynamics is given to a good approximation by Eqs. (\ref{eq:esegk}) and (\ref{eq:esegr}).
To understand the behavior near the scattering points we refer to Fig. \ref{fig:pocket2} and examine the vicinity
of the upper tip of the diamond pocket. There, the PDW mixes a state $|1\rangle = c^\dagger_{\bk,\sigma}|Fs\rangle$ on segment 1
with a state $|2\rangle = c^\dagger_{\bk+2\bQ_x,\sigma}|Fs\rangle$ on segment 4 via an intermediary state
$|3\rangle = c_{-\bk-\bQ_x,-\sigma}|Fs\rangle$. Here, $\sigma=\pm$ is the spin and $|Fs\rangle$
is the filled Fermi sea in the absence of the PDW. The reduced Hamiltonian within this subspace is
\begin{equation}
\label{eq:Hreduced}
\widetilde{H}=\left(\begin{array}{ccc}
\xi_{\bk+\frac{m}{\hbar}\vs} & 0 & \sigma e^{-i\theta^x}\Delta^x_{\bk+\frac{\bQ_x}{2}} \\
0 & \xi_{\bk+2\bQ_x+\frac{m}{\hbar}\vs} & \sigma e^{i\theta^x}\Delta^x_{\bk+\frac{3\bQ_x}{2}} \\
\sigma e^{i\theta^x}\Delta^x_{\bk+\frac{\bQ_x}{2}} & \sigma e^{-i\theta^x}\Delta^x_{\bk+\frac{3\bQ_x}{2}} & -\xi_{\bk+\bQ_x-\frac{m}{\hbar}\vs}
\end{array}
\right).
\end{equation}
By integrating out level $|3\rangle$ we obtain from it the effective low-energy Hamiltonian for the span of $\{|1\rangle$, $|2\rangle\}$
\begin{equation}
\label{eq:Heff}
H_{\rm eff}=\left(\begin{array}{cc}
\xi_1 & e^{-2i\theta^x}P \\
e^{2i\theta^x}P & \xi_2
\end{array}
\right),
\end{equation}
where
\begin{eqnarray}
\label{eq:xi12}
\xi_1(\bk,\br)&=&\xi_{\bk+\frac{m}{\hbar}\vs}+\frac{\left(\Delta^x_{\bk+\frac{\bQ_x}{2}}\right)^2}{\xi_{\bk+\bQ_x-\frac{m}{\hbar}\vs}},\\
\xi_2(\bk,\br)&=&\xi_{\bk+2\bQ_x+\frac{m}{\hbar}\vs}+\frac{\left(\Delta^x_{\bk+\frac{3\bQ_x}{2}}\right)^2}{\xi_{\bk+\bQ_x-\frac{m}{\hbar}\vs}},\\
P(\bk,\br)&=&\frac{\Delta^x_{\bk+\frac{\bQ_x}{2}}\Delta^x_{\bk+\frac{3\bQ_x}{2}}}{\xi_{\bk+\bQ_x-\frac{m}{\hbar}\vs}}.
\end{eqnarray}

We are interested in the eigenstate of $H_{\rm eff}$ that approaches $|1\rangle$ in region $a$ away from the tip and $|2\rangle$ in region $b$,
as defined in Fig. \ref{fig:pocket2}. The desired state is
\begin{equation}
\label{eq:tipstate}
|\nu\rangle =e^{i\chi}\left(
\begin{array}{c} e^{-2i\theta^x}\sqrt{\frac{1}{2}\left(1+\frac{\Delta\xi}{\Delta E}\right)} \\
\sqrt{\frac{1}{2}\left(1-\frac{\Delta\xi}{\Delta E}\right)}
\end{array}\right),
\end{equation}
where
\begin{eqnarray}
\label{eq:dxidef}
\Delta\xi&=&\xi_1-\xi_2,\\
\Delta E&=&\sqrt{\Delta\xi^2+4P^2},
\end{eqnarray}
and where the phase $\chi$ is to be chosen such that it approaches $2\theta^x$ for $\Delta\xi\gg |P|$, and 0 for $\Delta\xi\ll -|P|$.
For concreteness, we pick
\begin{equation}
\label{eq:chidef}
\chi(\bk,\br)=\left(1+\frac{\Delta\xi}{\Delta E}\right)\theta^x.
\end{equation}
More pertinent, from the perspective of elucidating the semiclassical dynamics, is the fact that the corresponding energy
\begin{equation}
\label{eq:tipenergy}
E=\frac{1}{2}(\xi_1+\xi_2+\Delta E),
\end{equation}
depends significantly on both $\bk_e$ and $\bk_h$ at the vicinity of the tip. Hence, while the $\bk$-space orbit essentially
coincides with constant $E(\bk_e)$ contours along the arcs of the diamond, this is no longer true near the scattering points.
We note that the situation is different from the one encountered for reconstruction scenarios due to particle-hole orders,
such as CDW.\cite{Gannot} There, at least in an ordered system, the $\bk$-space orbit is given precisely by the Fermi surface, thus
leading to the Onsager relation. We will not analyse in detail the semiclassical motion near the tip, but instead assume
that periodic solutions do exist, at least for a significant set of initial conditions. Our assumption is backed by numerical
calculations, which we detail below, that demonstrate quantum oscillations, albeit with broadened peaks and a slightly shifted
frequency compared to the $\vs=0$ case. We attribute this behavior to the spread in the semiclassical trajectories due to tip effects.

To proceed with semiclassical quantization of the periodic orbits we identify the conjugate momenta from the Lagrangian, Eq. (\ref{eq:L}),
\begin{eqnarray}
\label{eq:Pr}
\bp_{\br_c}&=&\eta\hbar\bk_c-\frac{e}{c}\tilde\bA(\br_c)+\eta\hbar\cbA_r(\bk_c,\br_c),\\
\bp_{\bk_c}&=&\eta\hbar\cbA_k(\bk_c,\br_c).
\end{eqnarray}
They enter the Bohr-Sommerfeld condition
\begin{equation}
\label{eq:BS1}
\frac{S}{\hbar}=\frac{1}{\hbar}\oint d\bk_c \!\cdot\! \bp_{\bk_c} + \frac{1}{\hbar}\oint d\br_c \!\cdot\! \bp_{\br_c}
=2\pi\left(n+\frac{\mu}{4}\right),
\end{equation}
where $n$ is integer and $\mu$ is the Maslov index of the trajectory. We first treat the ordered case $\theta^\alpha=0$, for which the Berry connections vanish owing to the fact that $|\nu\rangle$ is real, and
\begin{equation}
\label{eq:Sh1}
\frac{S}{\hbar}=\frac{1}{\hbar}\oint d\br_c \!\cdot\! \bp_{\br_c} =\oint d\br_c \!\cdot\! \left(\eta\bk_e-\frac{e}{\hbar c}\bA_{e\eta}\right).
\end{equation}
Assuming that no vortex cores are encountered and using Eq. (\ref{eq:esegr})
to express $d\br_c=l_B^2 d\bk_e\times\hat z$, the quantization rule reads
\begin{equation}
\label{eq:BS2}
\eta l_B^2 A_k+(\eta-1)\pi N_{\rm v}=2\pi\left(n+\frac{\mu}{4}\right),
\end{equation}
where $A_k$ is the area swept by the $\bk_e$ orbit and $N_{\rm v}$ is the number of vortices encircled.
Being a multiple of $2\pi$ the vortex contribution does not affect the quantization condition. The latter takes the
familiar Onsager form, except that $A_k$, as noted above, may deviate somewhat from the area of the pocket calculated for $\vs=0$.

\subsection{PDW phase disorder}

Before continuing with the analysis, let us note that the self-consistency argument of Section \ref{sub:wavepacket},
which favors constructing the wave packet such that $:\!\!\langle\Psi|\Psi^\dagger_\br\Psi_\br|\Psi\rangle\!\!:\,=\br_c$,
fails when $\theta^\alpha$ is not constant since then $\partial {\cal H}_\br/\partial\br_c\not \propto \sum_\ba \Psi^\dagger_{\br+\ba}\Psi_\br$.
Nevertheless, we assume that in the limit of $\Delta\rightarrow 0$, where the effects of the scattering events
are concentrated at small regions of phase space, this choice is still optimal. Taking this point of view, we
analyse the effects of PDW phase disorder within the same semiclassical framework used so far. We then contrast
its predictions with numerical calculations of the model in the next section.

In the presence of PDW phase disorder the Berry connections do not vanish. With the help of Eq. (\ref{eq:tipstate}) they evaluate
near the upper tip to
\begin{eqnarray}
\label{eq:Aktip}
\cbA_k&=&-\bnabla_{\bk_c}\chi,\\
\label{eq:Artip}
\cbA_r&=&-\bnabla_{\br_c}\chi+\left(1+\frac{\Delta\xi}{\Delta E}\right)\bnabla_{\br_c}\theta^x .
\end{eqnarray}
Their contribution to $S/\hbar$, coming from the orbit section that crosses the tip from
point $(\bk_i,\br_i)$ on segment 4 to point $(\bk_f,\br_f)$ on segment 1, is
\begin{eqnarray}
\nonumber
\label{eq:Berrycont}
&&-\eta\int_{\bk_i}^{\bk_f} d\bk_c\!\cdot\! \bnabla_{\bk_c}\left(\frac{\Delta\xi}{\Delta E}\right)\theta^x
-\eta\int_{\br_i}^{\br_f} d\br_c\!\cdot\! \bnabla_{\br_c}\left(\frac{\Delta\xi}{\Delta E}\right)\theta^x \\
&&=-\eta\int_{t_i}^{t_f} dt \frac{d}{dt}\left(\frac{\Delta\xi}{\Delta E}\right)\theta^x=-2\eta\theta^x(\br_1),
\end{eqnarray}
where $[t_i,t_f]$ is the time interval of the motion. The last equality is valid in the limit of vanishing PDW amplitude where
$(d/dt)(\Delta\xi/\Delta E)=2\delta(t_1)$, with $t_1$ the time at which the wave packet scatters at position $\br_c=\br_1$.

Collecting the contributions from the other scattering points yields $-2\eta[\theta^x(\br_1)-\theta^y(\br_2)-\theta^x(\br_3)+\theta^y(\br_4)]$.
Since the amplitude of the fundamental harmonic of the DOS is proportional to $\exp(iS/\hbar)$, its average over the fluctuations of the PDW
phases is suppressed by a Dingle factor.\cite{Gannot} In terms of the correlation lengths $\xi_{x,y}$ of $\theta^{x,y}$ it takes the form
\begin{equation}
\label{eq:Dingle}
R_D\sim e^{-4l_B^2(\Delta k_y/\xi_x+\Delta k_x/\xi_y)},
\end{equation}
where $\Delta_{k_{x,y}}$ are the $\bk$-space distances between the left-right and bottom-up scattering points, respectively.

We conclude this section by noting an interesting consequence of the fact that each scattering event adds $\pm 2\eta\theta^\alpha$
to $S/\hbar$. Counter to the general case, in the particular set of configurations where the PDW is disordered via $\pi$-phase
slips this contribution is a multiple of $2\pi$ and therefore has no effect on the DOS. Because of the reservation expressed at
the beginning of the discussion it is unclear to what extent this conclusion and Eq. (\ref{eq:Dingle}) should be trusted.
In the next section we provide numerical evidence, which show that although these results are not exact they hold to a
good degree of approximation.
While we do not have a microscopic reasoning for why disordering of the PDW should progress via creation of $\pi$-phase slips
we note that it offers a concrete model, which supports quantum oscillations in a highly disordered system without affecting
the frequency of the oscillations. This stands in contrast to first order scattering, e.g. from a CDW, where phase discommensurations
do not suppress the signal but leads to its appearance at higher harmonics.\cite{Gannot}

\begin{figure}[t!!!]
\centering
  \includegraphics[width=\linewidth,clip=true]{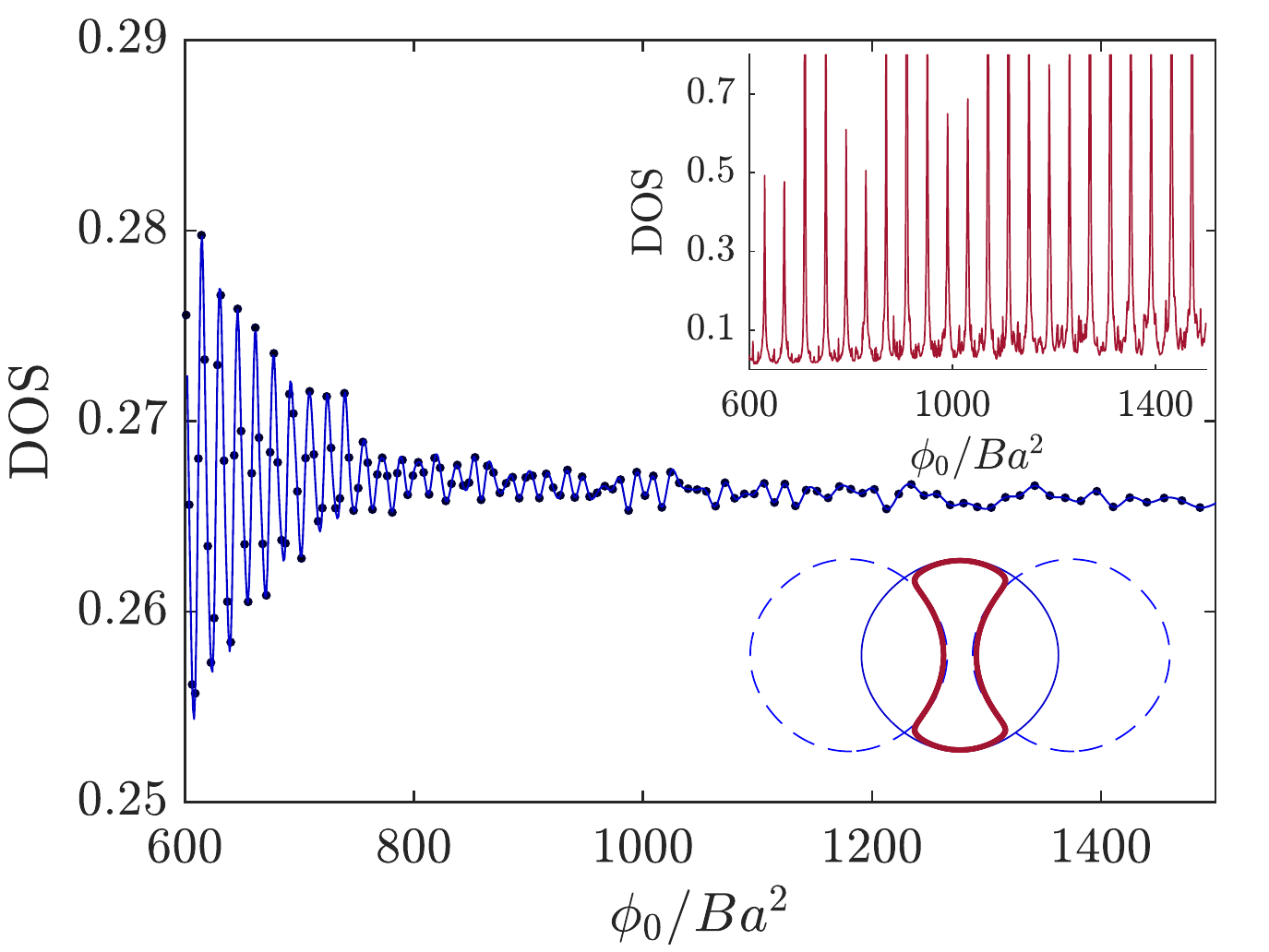}
  \caption{DOS as function of inverse magnetic field for a unidirectional $d$-wave PDW. First-order scattering leads at 
   $B=0$ to the red pocket, as indicated by the lower inset. Turning off $\vs$ results in the DOS oscillations depicted 
   in the upper inset, whose frequency is related by the Onsager rule to the area of the pocket. As shown by the main 
   figure these oscillations are eliminated once one includes the coupling to the superfluid velocity due to a vortex liquid. 
   New, faster oscillations appear at high fields, which are due to magnetic breakdown and correspond to an orbit around the 
   original Fermi surface, depicted in blue in the lower inset.}
  \label{fig:fo}
\end{figure}

\section{Numerical results}

\subsection{Model and method}

In order to establish the validity of the semiclassical results we use numerical methods to study the BdG Hamiltonian, 
Eq. (\ref{eq:H2}). We do so for the cases of a bidirectional $d$-wave PDW with $\Delta^x=\Delta^y=\Delta$, a unidirectional  
$d$-wave PDW with $\Delta^x=\Delta$ and $\Delta^y=0$, and for a unidirectional $s$-wave PDW, for which the pairing term in 
Hamiltonian (\ref{eq:H1}) reads $\Delta\sum_\br\cos(\bQ_x\cdot \br+\theta_\br)\Psi^\dagger_\br \Psi_\br$.

We carry out the calculations on a rectangular square lattice with $L_x$ sites and open boundary conditions in the $x$-direction. 
We apply periodic boundary conditions along the $y$-direction, such that $Q_y L_y=2\pi n$, where $L_y$ is the number of sites around 
the cylinder and $n$ is integer. To conform with the boundary conditions we use the Landau gauge $\bA=(0,B x,0)$.
The superfluid velocity field is constructed by summing over $N_{\rm v}=(2e/hc)BL_xL_y$ randomly placed vortex configurations, 
as described in Appendix \ref{App:vortex}. Note that the allowed magnetic fields in our finite lattice are constrained 
by the fact that $N_{\rm v}$ is integer.

Since quantum oscillations originate from the electron-like part of the DOS,\cite{Miller} our goal is to calculate 
\begin{equation}
\label{eq:elecDOS}
\rho(\omega)=-\frac{1}{\pi L_xL_y}{\rm Im\,}{\rm Tr}_e\left(\frac{1}{\omega-{\cal H}+i\delta}\right),
\end{equation}
where ${\rm Tr}_e$ is the partial trace over the electron-like part of $\Psi$. We concentrate on the $B$-dependence of the
zero-energy DOS, $\rho(0)$, and present this quantity in the following, while using $\delta=2.5\times 10^{-4}$. 
We have checked that the full DOS, defined by the trace over all the entries of $\Psi$,  
yields qualitatively similar results. In order to observe DOS oscillations one needs
to handle rather large systems. Typically, we use systems of size $L_x\times L_y=960\times 72$ and average over 2000-4000 
vortex liquid realizations. For such sizes calculation of $\rho(0)$ by straightforward diagonalization is tasking. Instead, 
we utilize the recursive Green's function method,\cite{recgreen} which allows to calculate the diagonal terms of the Green's function 
with computational cost that scales linearly with $L_x$.

\begin{figure}[t!!!]
\centering
  \includegraphics[width=\linewidth,clip=true]{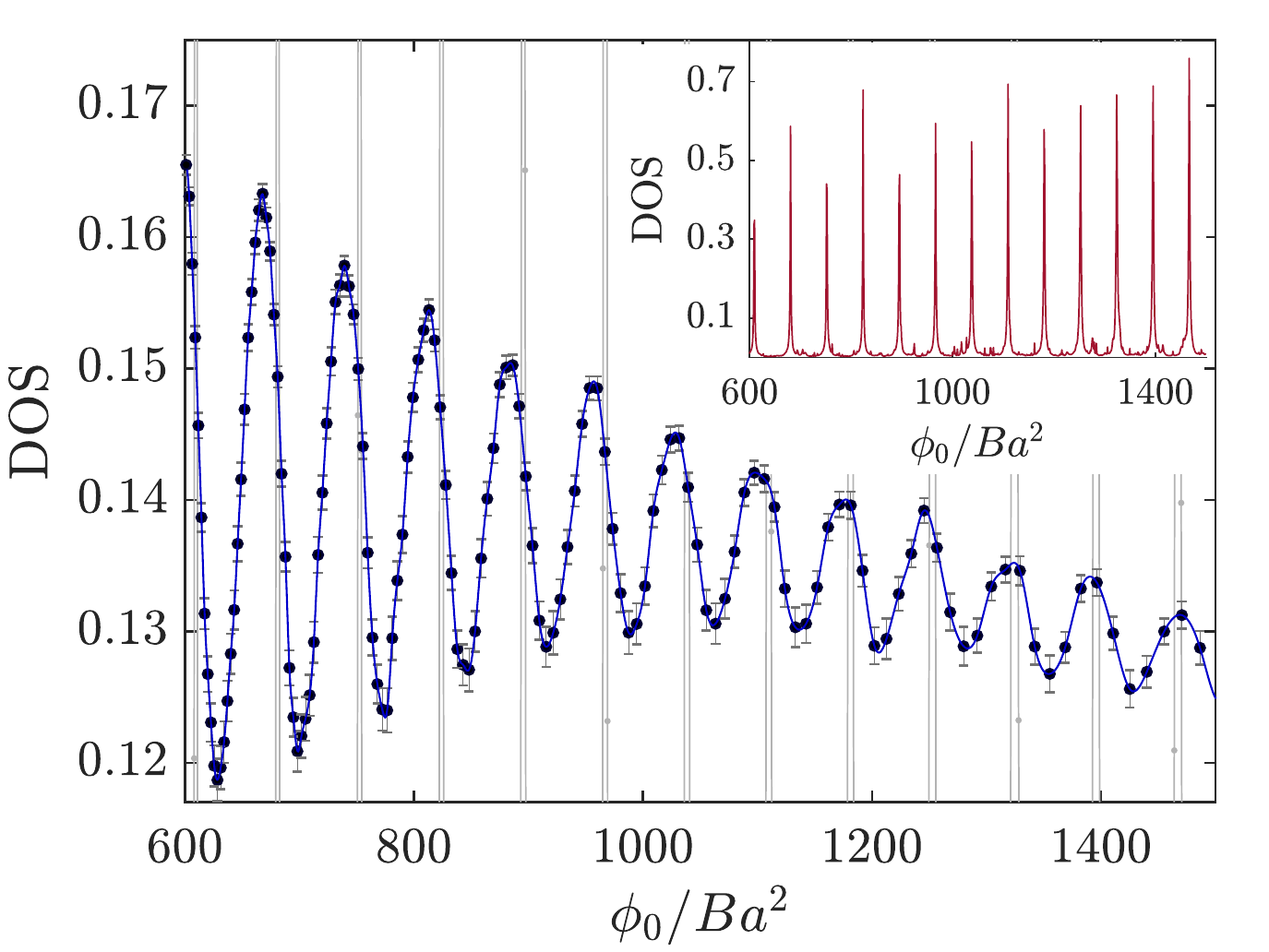}
  \caption{DOS as function of inverse magnetic field for a bidirectional $d$-wave PDW. Second-order scattering 
   results in diamond pockets, similar to those of Fig. \ref{fig:pocket2}. The DOS oscillations that originate from these pockets 
   for $\vs=0$ are depicted in the inset and serve as the background for the main figure. Their frequency follows the Onsager 
   relation. Turning on $\vs$ broadens the oscillations and very slightly shifts their frequency, but they are still clearly 
   visible in the main figure. }
  \label{fig:so-bi}
\end{figure}

\subsection{First-order scattering}

To check the ability of first order pockets to induce DOS oscillations we have considered a number of models where such pockets 
arise at zero magnetic field. A representative example is presented in Fig. \ref{fig:fo}, corresponding to a period 6 unidirectional 
$d$-wave PDW with $\Delta/t=0.4$ and $\mu/t=-3.2$. The lower inset depicts in red the pocket with area $A_k=0.025 A_{BZ}$, 
where $A_{BZ}$ is the area of the full Brillouin zone. The upper inset shows the DOS generated from Hamiltonian (\ref{eq:H2}) 
upon substituting $\vs=0$. Sharp oscillations are visible, at a frequency that follows the Onsager relation $F=(A_k/A_{BZ})\phi_0/a^2$, 
where $\phi_0=hc/e$. They originate from closed orbits around the pocket as depicted in Fig. \ref{fig:pocket1}.
We have found similar oscillations in all cases where first order pockets are present, as long as we set $\vs=0$.

Equally generic, is the disappearance of the oscillations once $\vs$ is included in the Hamiltonian, as demonstrated by the main panel 
of Fig. \ref{fig:fo}. We have found no exception to this statement in any system with first order pockets. Note that Fig. \ref{fig:fo} 
does show oscillations at high fields. However, they appear at a frequency $Fa^2/\phi_0=0.064$, which is very nearly the ratio between 
the area of the original Fermi surface and $A_{BZ}$. Hence, we associate the signal with a periodic orbit around the original Fermi surface 
due to magnetic breakdown. We have observed DOS oscillations from magnetic breakdown in many cases and found that they tend to commence 
at lower fields in the presence of vortices, as compared to the case $\vs=0$. 

\begin{figure}[t!!!]
\centering
  \includegraphics[width=\linewidth,clip=true]{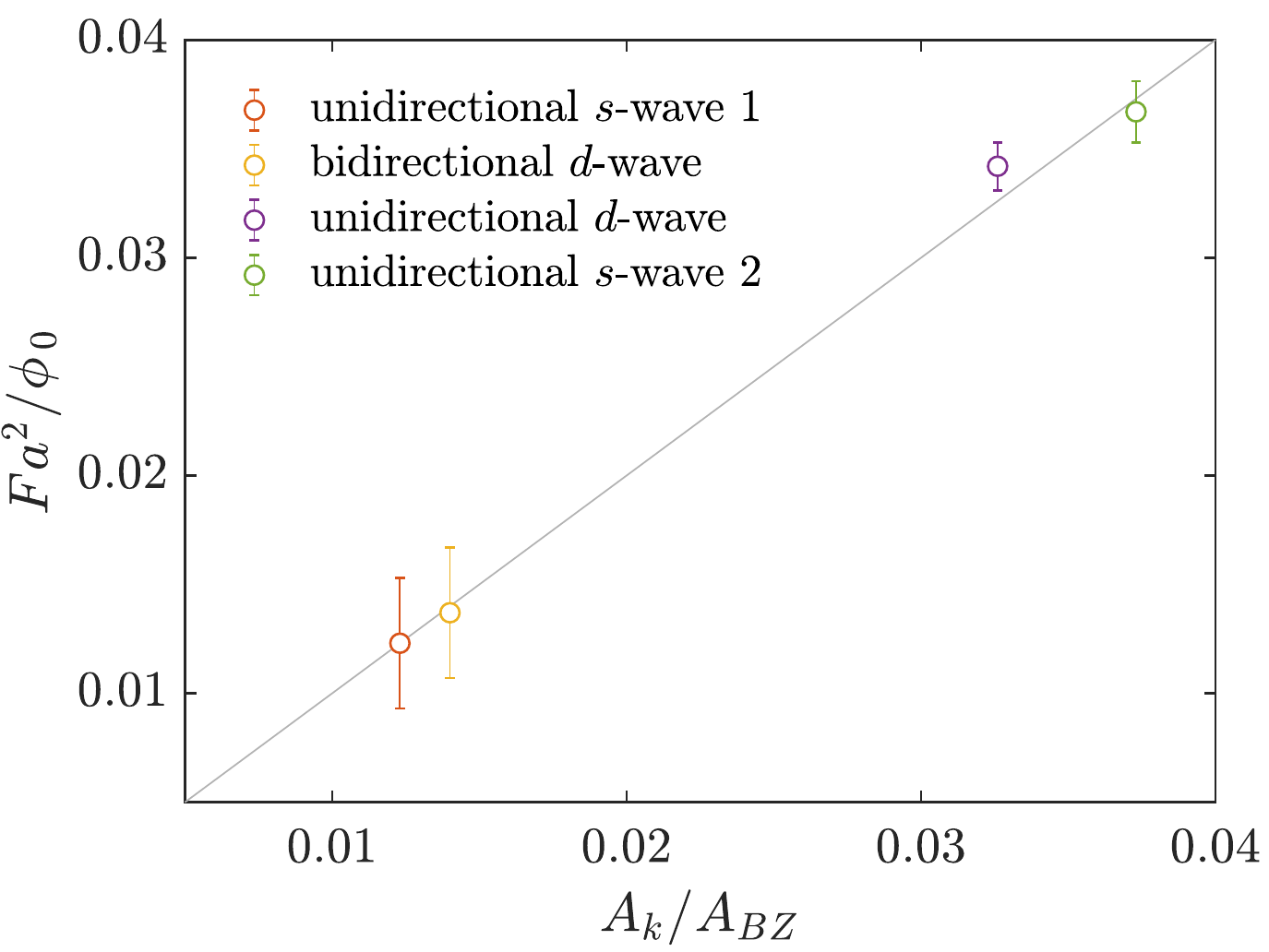}
  \caption{DOS oscillation frequency as function of the area of the second order pocket for various models
   that support such a pocket. The straight line depicts the Onsager relation.}
  \label{fig:freq}
\end{figure}

\subsection{Second-order pockets}

Similar to the situation with first-order pockets, second-order pockets also induce sharp DOS oscillations under the approximation $\vs=0$.
An example is shown in the inset of Fig. \ref{fig:so-bi}, which corresponds to a system hosting a period 6 bidirectional $d$-wave PDW 
with $\Delta/t=0.5$ and $\mu/t=-0.966$. The PDW generates a diamond-shaped pocket, as in Fig. \ref{fig:pocket2}, whose area is 
precisely related to the observed oscillation frequency via the Onsager rule. 

However, in contrast to first-order pockets, the oscillations stemming from second order pockets survive the inclusion of $\vs$, 
as illustrated by the main panel of Fig. \ref{fig:so-bi}. The DOS peaks are broader and slightly shifted compared to the signal 
for $\vs=0$. We attribute these effects to deviations of the semiclassical orbit from motion along constant energy 
contours near the tips of the pocket, as discussed in section \ref{section:so}. Nevertheless, the Onsager relation between the 
frequency of the DOS oscillations and the area of the pocket continues to hold to a good approximation. This fact is demonstrated by 
Fig. \ref{fig:freq} where we plot $F$ vs. $A_k$ for the model mentioned above, as well as for three other models. These include 
a period 8 unidirectional $d$-wave PDW with $\Delta/t=0.3$ and $\mu/t=-2.5$ and two period 8 unidirectional $s$-wave PDWs, one (1)
with $\Delta/t=0.4$ and $\mu/t=-3$, and the second (2) with $\Delta/t=0.25$  and $\mu/t=-2.5$. 

\begin{figure}[t!!!]
  \centering
  \includegraphics[width=\linewidth,clip=true]{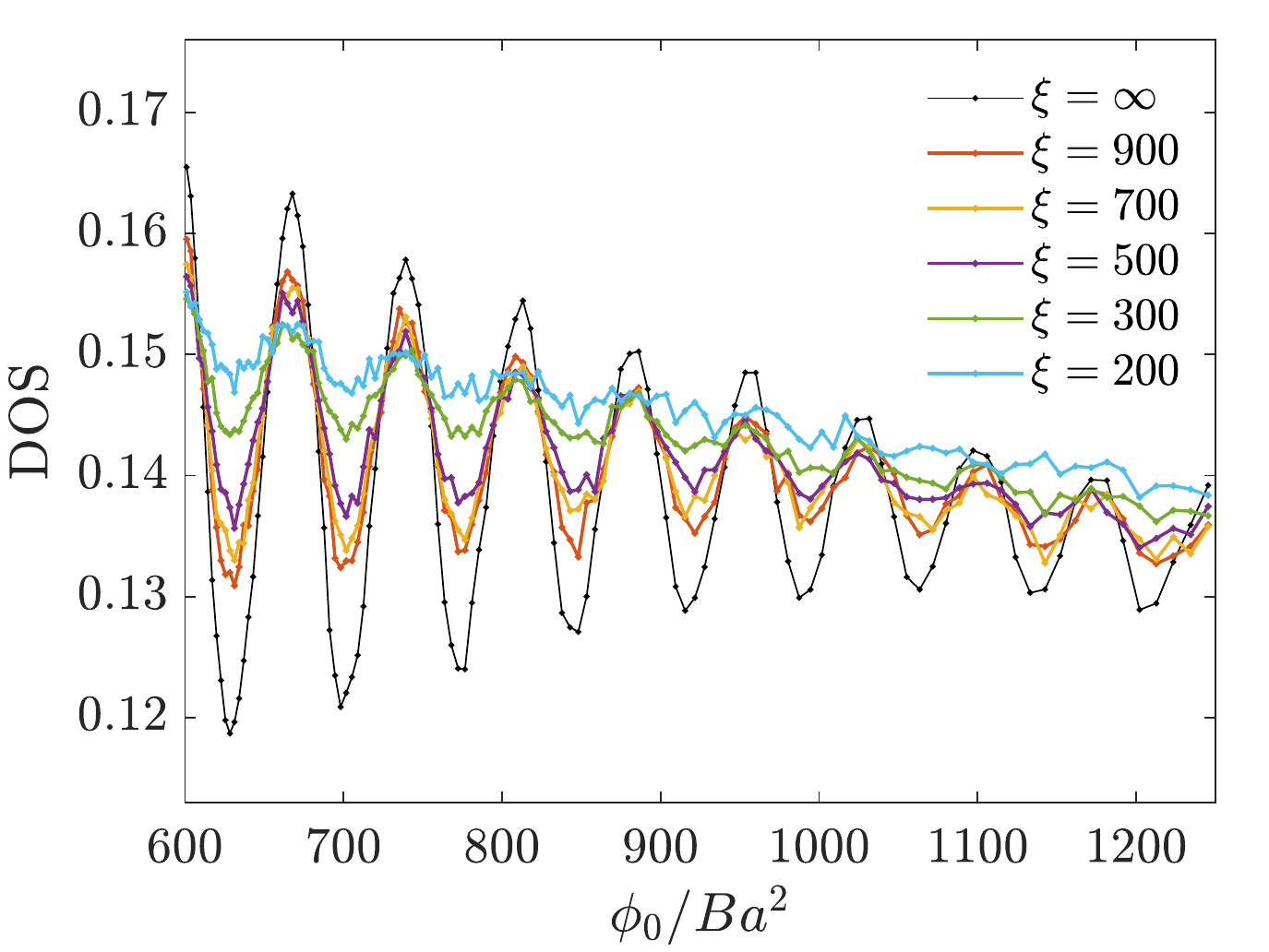}
  \caption{DOS oscillations for a system with a bidirectional $d$-wave PDW of the same parameters as in Fig. \ref{fig:so-bi}, 
   but whose phase is smoothly disordered in space with a correlation length $\xi$.}
  \label{fig:disorder-traces}
\end{figure}

\begin{figure}[t!!!]
  \centering
  \includegraphics[width=\linewidth,clip=true]{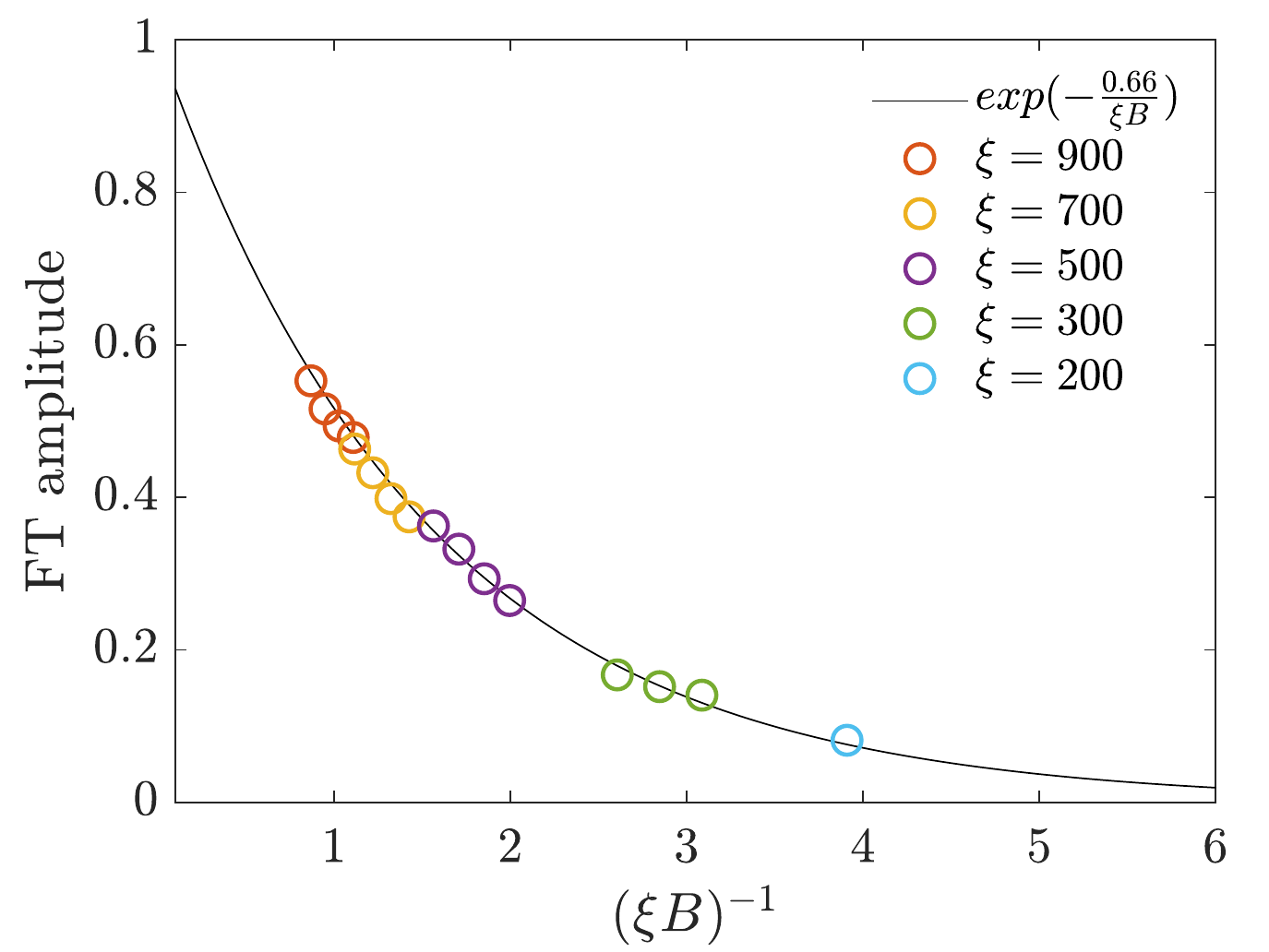}
  \caption{The dependence of the strength of the DOS oscillations, depicted in Fig. \ref{fig:disorder-traces}, on $(B\xi)^{-1}$. 
  The solid line is an exponential fit to the data.}
  \label{fig:disorder-decay}
\end{figure}

\subsection{PDW phase disorder}

Next, we consider the effects of phase disorder. We start by calculating the DOS for the bidirectional $d$-wave PDW, whose 
parameters are given above, in the case where its phases vary smoothly in space along a single direction, 
$\theta^\alpha_{\br+\ba/2}=\theta(x)$. We generate the phases using a one-dimensional random walk with a step size that is 
normally distributed. The standard deviation of the step is chosen to be $\sqrt{2/\xi}$, such that $\xi$ is the resulting 
correlation length for $\exp(i\theta)$. In addition, we smooth every phase configuration by convoluting it with a Gaussian 
of width 2. 

Fig. \ref{fig:disorder-traces} depicts the DOS for various values of $\xi$, where each trace is an average over 
2000-4000 realizations that differ both by their vortex distributions and by their PDW phase fields. The decay 
of the oscillations with increasing disorder is evident. In order to quantify it we split the $1/B$ range into 
sections, each containing 4 oscillations. We then plot, in Fig. \ref{fig:disorder-decay}, the amplitude of the 
Fourier transform peak resulting from each section after normalizing it by the corresponding amplitude for the clean system. 
We only present peaks that are significantly discernible from the background.
Based on our semiclassical result, Eq. (\ref{eq:Dingle}), we expect an exponential decay of the signal with $(\xi B)^{-1}$.
Note, however, that owing to the one-dimensional nature of the disorder that we use, only scattering events in the 
$x$-direction involve a phase difference and contribute to the Dingle factor. Consequently, the latter becomes 
$\exp(-4l_B^2\Delta k_y/\xi)$. For the diamond pocket created by the PDW considered here, this leads, in 
units where $\phi_0=1$, to $\exp(-0.584/\xi B)$. This is about 10\% slower than the best fit to the observed decay, depicted
by the solid line in Fig. \ref{fig:disorder-decay}.

\begin{figure}[t!!!]
  \centering
  \includegraphics[width=\linewidth,clip=true]{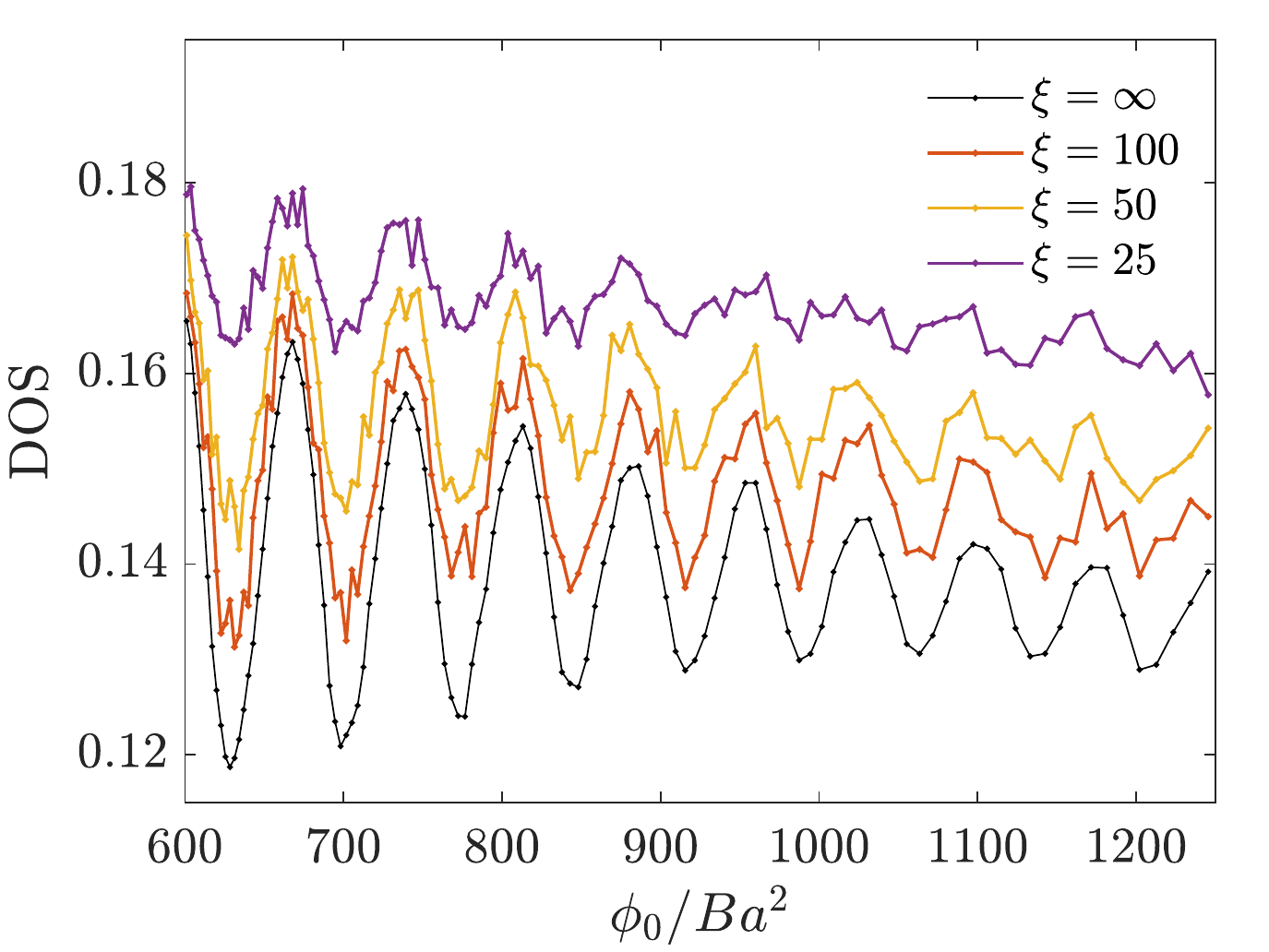}
  \caption{DOS oscillations for a system with a bidirectional $d$-wave PDW of the same parameters as in Fig. \ref{fig:so-bi},
   but whose phase is disordered via a random array of $\pi$-phase slips, resulting in a correlation length $\xi$.}
  \label{fig:phase-slips}
\end{figure}

Finally, we consider phase disorder in the form of a one-dimensional array of $\pi$-phase slips. The latter is generated
by a random walk, where a $\pi$ jump is introduced at each step with probability $(2\xi)^{-1}$. The resulting
$\theta^\alpha_{\br+\ba/2}=\theta(x)$ configuration is again smoothed out by convoluting it with a Gaussian of width 2.
The results for the DOS are presented in Fig. \ref{fig:phase-slips} and show that the quantum oscillations are considerably
more robust against disorder of this type, as expected on the basis of our semiclassical analysis.

\section{Discussion}

As alluded to in the Introduction, circumstantial evidence for a PDW state exists only in underdoped \BSCCOs (Bi2212). It takes the form of
bidirectional charge modulations around Abrikosov vortex cores with approximate periods $4a$ and $8a$, where $a$ is the lattice constant.
The appearance of two simultaneous CDWs with wave vectors $\bQ$ and $2\bQ$ is a natural consequence of $\bQ$-PDW modulations
existing in parallel to a uniform superconducting order.\cite{PDW-Senthil,PDW-Steve} In contrast, to date there are no measurements
showing quantum oscillations from a reconstructed Fermi surface in this compound. The opposite situation holds for underdoped
\YBCOs (YBCO), \YBCO248 and \HBCOs (Hg1201), where quantum oscillations from small Fermi pockets have been detected,\cite{PDW-review}
but there is no evidence to a PDW (very recently,\cite{Hsu} high magnetic field-resilient superconductivity has been observed in YBCO, 
which may involve a PDW). It is then natural to ask whether the observed oscillations in the latter group of materials can be
generated by a PDW, and what are the expected characteristics of PDW-induced quantum oscillations in Bi2212.

X-ray scattering has detected a bidirectional CDW in both YBCO and Hg1201.\cite{CDW-review} There is a significant correlation,
which follows the Onsager relation, between the area of the pocket obtained by folding the Fermi surface via the CDW wave vectors
and the frequency of the quantum oscillations.\cite{Tabis} However, the correlation length of the bidirectional
CDW in YBCO reaches only about 30 lattice constants,\cite{Gerber,Chang} while quantum oscillations commence at a magnetic field
where the cyclotron radius is approximately three times longer. The Dingle factor associated with such a ratio between the two length
scales would totally suppress the quantum oscillation signal.\cite{Gannot} This problem is even more pronounced in Hg1201.\cite{Tabis}
In YBCO, a CDW with the same period and considerably longer-ranged correlations appears under strong magnetic fields.\cite{Gerber,Chang}
Nevertheless, this CDW is unidirectional and it is not clear how it can give rise to pockets of the desired area. We have argued
that the longer-range CDW likely nucleates around vortices, where superconductivity is strongly suppressed, and is oriented due
to its Coulomb coupling to the chain layers.\cite{Yosef1,Yosef2} Here, we would like to raise the possibility that the unidirectional
CDW is actually a subsidiary order to a bidirectional PDW with twice the period, which competes with uniform superconductivity.
Such a PDW is capable of producing nodal "diamond" pockets via second order scattering, as shown in Fig. \ref{fig:pocket2}, that
match the observed oscillation frequencies. As mentioned, a PDW with wave vector $\bQ$ can also combine with uniform superconductivity
to produce a CDW at the same $\bQ$.
To the best of our knowledge, such a signal has not been observed thus far in YBCO,\cite{Chang-private}
but it may be too weak to detect at high fields
where the uniform order is effectively quenched. On the same note, one may contemplate the possibility that a bidirectional
PDW is also the parent order of the bidirectional CDW. However, the fact that experiments show no signatures of an additional
CDW with double the period under conditions where uniform superconductivity is still strong,\cite{Achkar} most likely rules it out.

\begin{figure}[t!!!]
  \centering
  \includegraphics[width=\linewidth,clip=true]{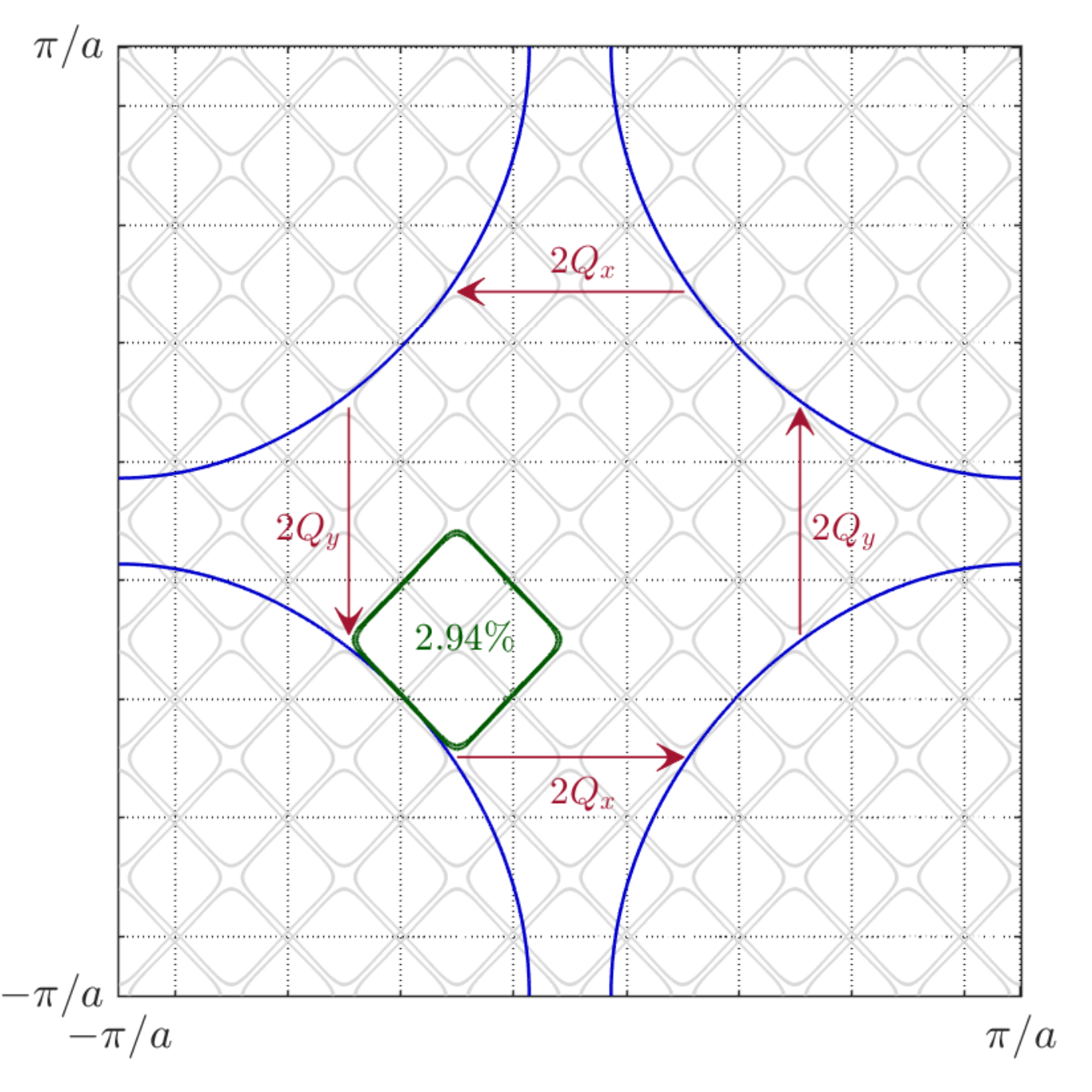}
  \caption{The expected electron pocket from a Fermi surface reconstruction in \BSCCOs by a PDW with the experimentally detected
  period.\cite{Halo-Edkins}}
  \label{fig:bisco}
\end{figure}

Applying the above scenario to Hg1201 would lead to the expectation that strong and ordered bidirectional PDW should appear around
vortex cores at higher fields than the highest field used so far ($\sim16$ T) but below the
onset field of quantum oscillations in this material ($\sim55$ T). Such a PDW would likely be accompanied by a CDW of
half the period (the same as of the observed low-field CDW), which is expected to be long-ranged and bidirectional, 
owing to the absence of orienting chains in the tetragonal Hg1201.


Finally, if a bidirectional PDW of period 8 exists in Bi2212, as the experiment suggests,\cite{Halo-Edkins} then it can lead to the 
formation of second-order pockets. In Fig. \ref{fig:bisco} we plot such pockets calculated for the case of a strong PDW $\Delta=0.5t$, 
and using the tight-binding dispersion of Ref. \onlinecite{He}. They are expected to induce DOS oscillations at a frequency of 
approximately 830T. We note, however, that if the PDW shares the very short correlation length of the observed CDWs within the 
halos then the oscillations would be massively damped. Furthermore, reducing the strength of the PDW in our calculations 
leads to small gaps that may be breached by magnetic breakdown and lead to different frequencies.

\acknowledgments
We thank S. Kivelson for useful comments. This research was supported by the Israel Science Foundation (Grant No. 701/17).

\appendix
\section{The spectrum of ${\cal H}$ with constant fields}
\label{App:diag}
Denote by $H(\bk)$ the Hamiltonian matrix for the case of constant fields in the representation where the
$\Lambda=\lambda_x\lambda_y$ Bragg vectors are ordered such that $\bq_i=-\bq_{\Lambda+1-i}$.
Its diagonalization, $H(\bk)=U_\bk^\dagger D_\bk U_\bk$, yields $2\Lambda$ eigenvalues
$E_{\bk 1}\leqslant ,\cdots,\leqslant E_{\bk \Lambda_\bk},-E_{\bk \Lambda_\bk+1}\geqslant ,\cdots,\geqslant -E_{\bk 2\Lambda}$,
with $E_{\bk n}\geqslant 0$.
We note that for $\vs\neq 0$ the number, $\Lambda_\bk$, of positive eigenvalues need not equal that of negative eigenvalues.
The eigenvectors form the columns of $U_\bk^\dagger$
\begin{eqnarray}
\label{eq:Udag}
\nonumber
U_\bk^\dagger&=&(|\nu_{\bk,1}\rangle,\cdots,|\nu_{\bk,2\Lambda}\rangle)\\
&=&\left[\left(\begin{array}{c} u_{\bk+\bq_1,1} \\ v_{\bk+\bq_1,1} \\ \vdots \\ u_{\bk+\bq_\Lambda,1} \\ v_{\bk+\bq_\Lambda,1}
\end{array}\right) \cdots
\left(\begin{array}{c} u_{\bk+\bq_1,2\Lambda} \\ v_{\bk+\bq_1,2\Lambda} \\ \vdots \\ u_{\bk+\bq_\Lambda,2\Lambda} \\ v_{\bk+\bq_\Lambda,2\Lambda}
\end{array}\right)\right].
\end{eqnarray}

The spectra of $H(\bk)$ and $H(-\bk)$ are related by symmetry. The blocks $H_{\bq\bq'}(\bk)$, Eq. (\ref{eq:Hk}),
obey the relation $H^*_{-\bq-\bq'}(-\bk)=-\tau_2 H_{\bq\bq'}(\bk)\tau_2$, from which it follows that
\begin{equation}
\label{eq:Hksymm}
T_2 H^*(-\bk)= -H(\bk)T_2,
\end{equation}
where
\begin{equation}
\label{eq:Sigma2}
T_2=\left(\begin{array}{ccc} 0 & & -i\tau_2 \\ & \reflectbox{$\ddots$} & \\ -i\tau_2 & & 0 \end{array}\right).
\end{equation}
Given an eigenvector $|\nu_{-\bk,n}\rangle$ of $H(-\bk)$ corresponding to an eigenvalue $\cal E$ (positive or negative)
we can apply Eq. (\ref{eq:Hksymm}) to $|\nu_{-\bk,n}\rangle^*$ and find that $T_2|\nu_{-\bk,n}\rangle^*$
is an eigenvector of $H(\bk)$ with energy $-\cal E$. Hence, we conclude that for non-degenerate ${\cal E}>0$ and up to
a phase $T_2|\nu_{-\bk,n}\rangle^*=|\nu_{\bk,\Lambda_\bk+n}\rangle$. Similarly,
$T_2|\nu_{-\bk,\Lambda_{-\bk}+n}\rangle^*=|\nu_{\bk,n}\rangle$, for non-degenerate ${\cal E}<0$. By proper orthogonalization
these statements can be made true also for the degenerate case.

Guided by the above observation we define the quasiparticle creation operators, for $n=1,\cdots,\Lambda_\bk$, as follows
\begin{eqnarray}
\label{eq:gammadefp}
\gamma^\dagger_{\bk n+}&=&\sum_\bq \Psi_{\bk+\bq}^\dagger\left(\begin{array}{c} u_{\bk+\bq,n} \\ v_{\bk+\bq,n} \end{array}\right),\\
\nonumber
\label{eq:gammadefm}
\gamma^\dagger_{\bk n-}&=&\sum_\bq \left( u^*_{-\bk+\bq,\Lambda_{-\bk}+n},v^*_{-\bk+\bq,\Lambda_{-\bk}+n}\right) \Psi_{-\bk+\bq}\\
&=&\sum_\bq \left( u_{\bk+\bq,n},v_{\bk+\bq,n}\right)(-i\tau_2) \Psi_{-\bk-\bq}.
\end{eqnarray}
Using the fact that the columns and rows of $U_\bk^\dagger$ form an orthonormal basis one can verify that
$\{\gamma_{\bk n\eta},\gamma_{\bk' n'\eta'}\}=0$ and
$\{\gamma_{\bk n\eta},\gamma^\dagger_{\bk' n'\eta'}\}=\delta_{\bk\bk'}\delta_{nn'}\delta_{\eta\eta'}$, where $\eta=\pm$.

\vspace{0.1cm}
The BdG Hamiltonian is diagonal in terms of these operators
\begin{equation}
\label{eq:Hdiag}
{\cal H}=\sum_\bk\sum_{n=1}^{\Lambda_\bk}E_{\bk n}\left(\gamma_{\bk n+}^\dagger \gamma_{\bk n+} - \gamma_{\bk n-}
\gamma_{\bk n-}^\dagger\right),
\end{equation}
with the $\gamma$-vacuum, $|g\rangle$, as a ground state and quasiparticle Bloch states
\begin{equation}
\label{eq:qpstates}
\gamma^\dagger_{\bk n\pm}|g\rangle=\frac{\Lambda^{1/2}}{L}
\sum_\br e^{i(\bk\pm\frac{m}{\hbar}\vs \mp\frac{e}{\hbar c}\bA)\cdot\br}\left\{
\begin{array}{c} \Psi_\br^\dagger\varphi_{\bk n+}(\br)|g\rangle \\
\varphi^\dagger_{\bk n-}(\br)\Psi_\br |g\rangle \end{array},\right. \\
\end{equation}
whose periodic parts are given by
\begin{equation}
\label{eq:phis}
\varphi^\dagger_{\bk n\pm}(\br)=\frac{1}{\Lambda^{1/2}}\sum_\bq e^{\mp i\bq\cdot \br}
\left\{\begin{array}{c} (u^*_{\bk+\bq n},v^*_{\bk+\bq n})\\
(v_{\bk+\bq n},-u_{\bk+\bq n}) \end{array}.\right.
\end{equation}

\section{Expectation values of $|\Psi\rangle$}
\label{App:expect}

In order to calculate $:\!\!\langle\Psi|\hat\br|\Psi\rangle\!\!:$ for the position operator,
 $\hat\br=\sum_\br \br\Psi^\dagger_\br \Psi_\br$, we use Eqs. (\ref{eq:gammadefp}) and (\ref{eq:phis}) to find
\begin{eqnarray}
\label{eq:rp1}
\nonumber
\!\!\!\!\!\!\!\!\!\!\!\!\langle g|\gamma_{\bk' \eta}\Psi^\dagger_\br\Psi_\br\gamma^\dagger_{\bk \eta}|g \rangle&=&
\langle g|\Psi^\dagger_\br\Psi_\br |g\rangle\delta_{\bk\bk'}\\
\!\!\!\!\!\!\!\!\!\!\!\!&+&\eta\frac{\Lambda}{L^2}e^{i(\bk-\bk')\cdot\br}\varphi^\dagger_{\bk'+}(\br)\varphi_{\bk+}(\br),
\end{eqnarray}
where $\eta=\pm$. Combining this result with Eq. (\ref{eq:wavepacket}) gives
\begin{eqnarray}
\label{eq:rp2}
\nonumber
\!\!\!\!\!\! :\!\!\langle\Psi_\eta|\hat\br|\Psi_\eta\rangle\!\!:&=&i\eta\sum_{\bk,\bk'}\delta_{\bk\bk'}
\bnabla_\bk{\Big[}W(\bk'-\bk_c)W(\bk-\bk_c) \\
\!\!\!\!\!\! &\times& e^{i\eta(\bk'-\bk)\cdot(\br_c-\cbA_k)}\langle\varphi_{\bk'+}|\varphi_{\bk+}\rangle{\Big]}=\br_c,
\end{eqnarray}
where we have used $\br e^{i\bk\cdot\br}=-i\bnabla_\bk e^{i\bk\cdot\br}$ and integrated by parts over $\bk$ to arrive
at the first equality. The final result then follows from the periodicity of $W$ and the definition of $\cbA_k$.

For the purpose of evaluating the semiclassical Lagrangian we are in need of $:\!\!\langle\Psi_\eta|i\frac{d}{dt}|\Psi_\eta\rangle\!\!:$.
Here, we draw attention to the contribution originating from the implicit time dependence of $\gamma^\dagger_{\bk\eta}$
\begin{eqnarray}
\label{eq:ddt}
\nonumber
&&:\!\!\langle\Psi_\pm|\frac{\Lambda^{1/2}}{L}\sum_\br
e^{i[\bk\pm(\frac{m}{\hbar}\vs -\frac{e}{\hbar c}\bA)]\cdot\br} \\
\nonumber
&&\times\left[\pm\frac{d}{dt}\left(\frac{e}{\hbar c}\bA-\frac{m}{\hbar}\vs\right)\cdot\br+\dot{\br}_c\cdot\bnabla_{\br_c}\right]
\left\{
\begin{array}{c} \Psi_\br^\dagger\varphi_{\bk +}(\br)|g\rangle\!\!: \\
\varphi^\dagger_{\bk -}(\br)\Psi_\br |g\rangle\!\!: \end{array}.\right. \\
\end{eqnarray}
Using similar steps to those leading to Eq. (\ref{eq:rp2}) one finds that the first term evaluates to
$\br_c\cdot(d/dt)[(e/\hbar c)\bA-(m/\hbar)\vs]$, while the second gives rise to the Berry connection $\cbA_r$. Adding the
remaining contributions and a total time derivative, which does not affect the equations of motion, results in Eq. (\ref{eq:L}).

\vspace{0.3cm}
\section{Construction of the vortex liquid configuration}
\label{App:vortex}
Constructing a vortex configuration on the cylinder requires a smooth phase-gradient field whose circulation vanishes around every 
plaquette, except the one containing the core. To this end, we start by considering a $2L_x\times 10^4L_y$ 
lattice with open boundary conditions, whose geometric center is defined as the origin. The phase on each site, $\phi_\br$, 
takes the value of the site's azimuthal angle. To keep the phase gradient smooth it is defined by 
\begin{equation}
\label{eq:nablaphi}
\bnabla\phi_{\br+\ba/2}=\phi_{\br+\ba}-\phi_\br-2\pi\Theta(\phi_{\br+\ba}-\phi_\br-\pi){\rm sgn}(\phi_{\br+\ba}-\phi_\br),
\end{equation}
where $\Theta$ is the step function. Next, we sum over 1000 replicas of this configuration, each with a vortex center that is
successively translated by $10L_y$ along the $y$-direction. The configuration obtained by closing the central
$L_y$ strip on itself fulfills the above requirements, up to a very small flux on plaquettes crossing the seam.
The vortex liquid is obtained by superposing $N_{\rm v}$ such vortex configurations with randomly placed cores.
It is used to calculate the $\vs$ field via ${{\vs}_{}}_{\br+\ba/2}=(\hbar/2m)[{\bm \nabla}\phi_{\br+\ba/2}+(2e/\hbar c)\bA_{\br+\ba/2}]$.
Finally, to avoid any residual total current through the system we subtract the spatial average of the velocity from every $y$-bond.


\begin{thebibliography}{999}

\bibitem{intertwined} E.~Fradkin, S.~A.~Kivelson, and J.~M.~Tranquada,
\rmp {\bf 87}, 457 (2015).

\bibitem{CDW-review} R.~Comin and A.~Damascelli,
Annu. Rev. Condens. Matter Phys. {\bf 7}, 369 (2016)

\bibitem{PDW-review} D.~F.~Agterberg, J.~C.~S.~Davis, S.~D.~Edkins, E.~Fradkin, D.~J.~Van Harlingen, S.~A.~Kivelson,P.~A.~Lee,
L.~Radzihovsky, J.~M.~Tranquada, and Y.~Wang,
Annu. Rev. Condens. Matter Phys. {\bf 11}, 231 (2020).

\bibitem{FF} P.~Fulde and R.~A.~Ferrell,
Phys. Rev. {\bf 135}, A550 (1964)

\bibitem{LO} A.~I.~Larkin and Yu.~N.~Ovchinnikov,
Sov. Phys. JETP {\bf 20}, 762 (1965).

\bibitem{PDW-Himeda} A.~Himeda, T.~Kato, and M.~Ogata,
\prl {\bf 88}, 117001 (2002).

\bibitem{PDW-Berg} E.~Berg, E.~Fradkin, E.-A.~Kim, S.~A.~Kivelson, V.~Oganesyan, J.~M.~Tranquada, and S.~C.~Zhang,
\prl {\bf 99}, 127003 (2007).

\bibitem{PDW-Lee} P.~A.~Lee,
Phys. Rev. X {\bf 4}, 031017 (2014).

\bibitem{PDW-2d-tJ} M.~Raczkowski, M.~Capello, D.~Poilblanc, R.~Fr${\rm \acute{e}}$sard, and A.~M.~Ole${\rm \acute{s}}$,
\prb {\bf 76}, 140505(R) (2007).

\bibitem{PDW-Yang} K-Y.~Yang, W.~Q.~Chen2, T.~M.~Rice, M.~Sigrist and F.-C.~Zhang,
New J. Phys. {\bf 11}, 055053 (2009).

\bibitem{Kopp} F.~Loder, A.~P.~Kampf, and T.~Kopp,
\prb {\bf 81}, 020511(R)(2010).

\bibitem{Corboz} P.~Corboz, T.~M.~Rice, and M.~Troyer,
\prl {\bf 113}, 046402 (2014).

\bibitem{Halo-Edkins} S.~D.~Edkins A.~Kostin, K.~Fujita, A.~P.~Mackenzie, H.~Eisaki, S.~Uchida, S.~Sachdev, M.~J.~Lawler,
E.-A.~Kim, J.~C.~S.~Davis and M.~H.~Hamidian,
{\it Science} {\bf 364}, 976 (2019).

\bibitem{Sebastian-review} S.~E.~Sebastian and C.~Proust,
Annu. Rev. Condens. Matter Phys. {\bf 6}, 411 (2015).

\bibitem{H-S} N.~Harrison and S.~E.~Sebastian,
\prl {\bf 106}, 226402 (2011).

\bibitem{Sachdev} A.~Allais, D.~Chowdhury, and S.~Sachdev,
\natcomm {\bf 5}, 5771 (2014).

\bibitem{Gannot} Y.~Gannot, B.~J.~Ramshaw, and S.~A.~Kivelson,
\prb {\bf 100}, 045128 (2019).

\bibitem{Norman} A.~J.~Millis and M.~R.~Norman, 
\prb {\bf 76}, 220503(R) (2007).

\bibitem{Steve-uni} H.~Yao, D.-H.~Lee, and S.~Kivelson,
\prb {\bf 84}, 012507 (2011).

\bibitem{Mohit} S.~Banerjee, S.~Zhang, and M.~Randeria,
\natcomm {\bf 4}, 1700 (2013).

\bibitem{DDW-Lee} K.-T. Chen and P. A. Lee,
\prb {\bf 79}, 180510 (2009).

\bibitem{DDW-Chakravarty} Z.~Wang and S.~Chakravarty,
\prb {\bf 93}, 184505 (2016).

\bibitem{loop-Senthil} A.~Allais and T.~Senthil, 
\prb {\bf 86}, 045118 (2012).

\bibitem{loop-Vafek} L.~Wang and O.~Vafek, 
\prb {\bf 88}, 024506 (2013).

\bibitem{Shirit} S.~Baruch and D.~Orgad,
\prb {\bf 77}, 174502 (2008).

\bibitem{Zelli-QO} M.~Zelli, C.~Kallin, and A.~J.~Berlinsky,
\prb {\bf 86}, 104507 (2012).

\bibitem{Davis-Norman} M.~R.~Norman and J.~C.~S.~Davis,
Proc. Natl. Acad. Sci. {\bf 115}, 5389 (2018).

\bibitem{PDW-Senthil} Z.~Dai, Y.-H.~Zhang, T.~Senthil, and P.~A.~Lee,
\prb {\bf 97}, 174511 (2018).

\bibitem{Anderson} P.~W.~Anderson,
arXiv:cond-mat/9812063.

\bibitem{Niu-semi} G.~Sundaram and Q.~Niu,
\prb {\bf 59}, 14915 (1999).

\bibitem{Liang-semi} L.~Liang, S.~Peotta, A.~Harju, and P.~T${\rm \ddot{o}}$rm${\rm \ddot{a}}$,
\prb {\bf 96}, 064511 (2017).

\bibitem{Niu2020} Z.~Wang, L.~Dong, C.~Xiao, and Q.~Niu,
arXiv:2008.11374.

\bibitem{Miller} P.~Miller and B.L.~Gyorffy,
J. Phys. Condens. Matter {\bf 7}, 5579 (1995).

\bibitem{recgreen} P.~A.~Lee and D.~S.~Fisher,
\prl {\bf 47}, 882 (1981).

\bibitem{PDW-Steve} Y.~Wang, S.~D.~Edkins, M.~H.~Hamidian, J.~C.~S.~Davis, E.~Fradkin, and S.~A.~Kivelson,
\prb {\bf 97}, 174510 (2018).

\bibitem{Hsu} Y.-T.~Hsu, M.~Hartstein, A.~J.~Davies, A.~J.~Hickey, M.~K.~Chan, J.~Porras, 
T.~Loew, S.~V.~Taylor, H.~Liu, A.~G.~Eaton, M.~Le~Tacon, H.~Zuo, J.~Wang, Z.~Zhu, G.~G.~Lonzarich, 
B.~Keimer, N.~Harrison, and S.~E.~Sebastian,
Proc. Natl. Acad. Sci. {\bf 118}, e2021216118 (2021).

\bibitem{Tabis} W.~Tabis, B.~Yu, I.~Bialo, M.~Bluschke, T.~Kolodziej, A.~Kozlowski, E.~Blackburn, K.~Sen, E.~M.~Forgan,
M.~v.~ Zimmermann, Y.~Tang, E.~Weschke, B.~Vignolle, M.~Hepting, H.~Gretarsson, R.~Sutarto, F.~He, M.~Le~Tacon,
N.~Bari${\rm \check{s}}$i${\rm \acute{c}}$, G.~Yu, and M.~Greven,
\prb{\bf 96}, 134510 (2017).

\bibitem{Gerber} S.~Gerber, H.~Jang, H.~Nojiri, S.~Matsuzawa, H.~Yasumura, D.~A.~Bonn, R.~Liang, W.~N.~Hardy, Z.~Islam, A.~Mehta,
S.~Song, M.~Sikorski, D.~Stefanescu, Y.~Feng, S.~A.~Kivelson, T.~P.~Devereaux, Z.-X.~Shen, C.-C.~Kao, W.-S.~Lee, D.~Zhu, and J.-S.~Lee,
{\it Science} {\bf 350}, 949 (2015).

\bibitem{Chang} J.~Chang, E.~Blackburn, O.~Ivashko, A.~T.~Holmes, N.~B.~Christensen, M.~H${\rm \ddot{u}}$cker, R.~Liang, D.~A.~Bonn,
W.~N.~Hardy, U.~R${\rm \ddot{u}}$tt, M.~v.~Zimmermann, E.~M.~Forgan, and S.~M.~Hayden,
Nat. Commun. {\bf 7}, 11494 (2016).

\bibitem{Yosef1} Y.~Caplan, G.~Wachtel and D.~Orgad,
\prb {\bf 92}, 224504 (2015).

\bibitem{Yosef2} Y.~Caplan and D.~Orgad,
\prl {\bf 119}, 107002 (2017).

\bibitem{Chang-private} E.~Blackburn \etal, in preparation; M.~Bluschke, private communication.

\bibitem{Achkar} A.~J.~Achkar, X.~Mao, C.~McMahon, R.~Sutarto, F.~He, R.~Liang, D.~A.~Bonn, W.~N.~Hardy, and D.~G.~Hawthorn,
\prl {\bf 113}, 107002 (2014).

\bibitem{He} R.-H.~He, M.~Hashimoto, H.~Karapetyan, J.~D.~Koralek, J.~P.~Hinton, J.~P.~Testaud, V.~Nathan, Y.~Yoshida, H.~Yao,
K.~Tanaka, W.~Meevasana, R.~G.~Moore, D.~H.~Lu, S.-K.~Mo, M.~Ishikado, H.~Eisaki, Z.~Hussain, T.~P.~Devereaux, S.~A.~Kivelson, 
J.~Orenstein, A.~Kapitulnik, and Z.-X. Shen,
{\it Science} {\bf 331}, 1579 (2011).

\end{thebibliography}
\end{document}